%% file: main.tex
\documentclass[10pt,twocolumn,letterpaper]{article}

\usepackage[pagenumbers]{cvpr} 

\input{preamble}

\definecolor{cvprblue}{rgb}{0.21,0.49,0.74}
\usepackage[pagebackref,breaklinks,colorlinks,citecolor=cvprblue]{hyperref}
\usepackage{algorithm}
\usepackage{algpseudocode}
\usepackage{multirow}
\usepackage{tabularx}
\usepackage{caption}
\usepackage{subcaption}
\usepackage{bm}
\DeclareMathOperator*{\argmax}{arg\,max}
\DeclareMathOperator*{\argmin}{arg\,min}
\def\paperID{13781} 
\def\confName{CVPR}
\def\confYear{2024}

\title{VA3: Virtually Assured Amplification Attack on Probabilistic Copyright Protection for Text-to-Image Generative Models}

\author{Xiang Li\footnotemark[1] \qquad  Qianli Shen\footnotemark[1] \qquad Kenji Kawaguchi \\
National University of Singapore\\
{\tt\small \{xiangli,qianli,kenji\}@comp.nus.edu.sg}
}

\begin{document}
\twocolumn[{%
\renewcommand\twocolumn[1][]{#1}%
\maketitle
\begin{center}
    \centering
    \captionsetup{type=figure}
    \includegraphics[width=\textwidth]{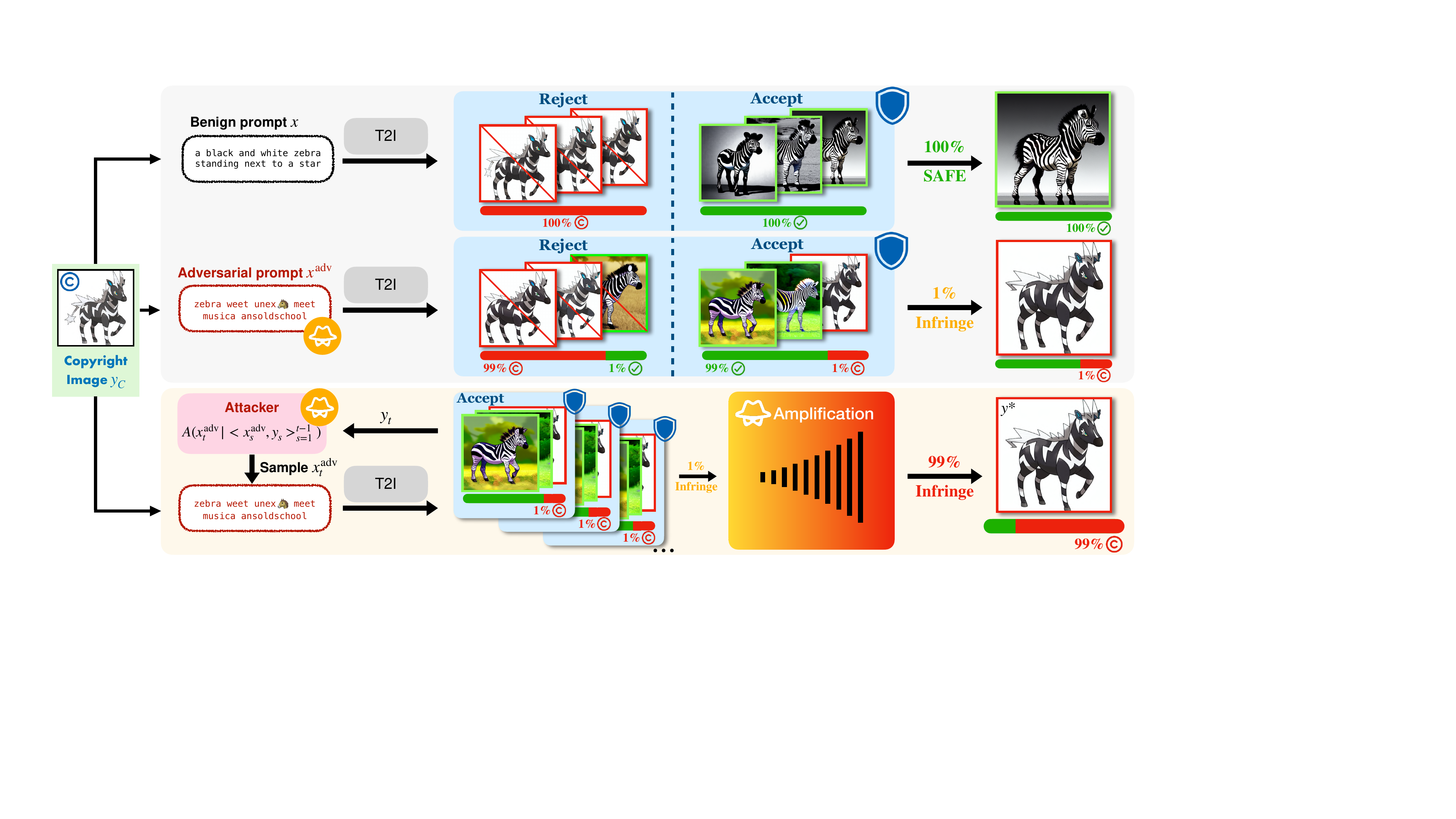}
    \captionof{figure}{Given a copyrighted image $y_C$ and a Text-to-Image (T2I) generative model with probabilistic copyright protection, our proposed virtually assured amplification attack (VA3) significantly amplifies the probability of producing infringing generations with persistent interactions of online adversarial prompt selection.}
    \label{fig:demo}
\end{center}%
}]

\renewcommand{\thefootnote}{\fnsymbol{footnote}}
\footnotetext[1]{Equal contribution}

\renewcommand{\thefootnote}{\arabic{footnote}}

\input{sec/0_abstract}    
\input{sec/1_intro}
\input{sec/2_related_work}
\input{sec/3_preliminary}

\input{sec/4_method}

\input{sec/5_exp}

\input{sec/6_conclusion}
\input{sec/7_acknowledgement}

{
    \small
    \bibliographystyle{ieeenat_fullname}
    \bibliography{main}
}

\input{sec/X_suppl}

\end{document}

%% file: preamble.tex
\usepackage[dvipsnames]{xcolor}

\newcommand{\nb}[3]{\ifthenelse{\boolean{include-notes}}{{\colorbox{#2}{\bfseries\sffamily\scriptsize\textcolor{white}{#1}}}{\ \textcolor{#2}{\sf\small\textit{#3}}}}{}}

\usepackage{mathtools}
\usepackage{amsthm}
\theoremstyle{plain}
\newcommand{\Ourmethod}{Amplification\xspace} 
\newcommand{\ourmethod}{amplification\xspace}
\newcommand{\ourmethodshort}{Amp.\xspace}
\newtheorem{theorem}{Theorem}
\newtheorem{theorem_X}{Theorem}
\newtheorem{definition}{Definition}

%% file: sec/0_abstract.tex
\begin{abstract}
 The booming use of text-to-image generative models has raised concerns about their high risk of producing copyright-infringing content.
 While probabilistic copyright protection methods provide a probabilistic guarantee against such infringement, in this paper, we introduce Virtually Assured Amplification Attack (VA3), a novel online attack framework that exposes the vulnerabilities of these protection mechanisms.
 The proposed framework significantly amplifies the probability of generating infringing content on the sustained interactions with generative models and a non-trivial lower-bound on the success probability of each engagement.
 Our theoretical and experimental results demonstrate the effectiveness of our approach under various scenarios.
 These findings highlight the potential risk of implementing probabilistic copyright protection in practical applications of text-to-image generative models.  Code is available at \href{https://github.com/South7X/VA3}{https://github.com/South7X/VA3}.
\end{abstract}


%% file: sec/1_intro.tex
\section{Introduction}
\label{sec:intro}


\begin{figure*}[t]
    \centering
    \begin{subfigure}[t]{0.24\textwidth}
        \includegraphics[width=\textwidth]{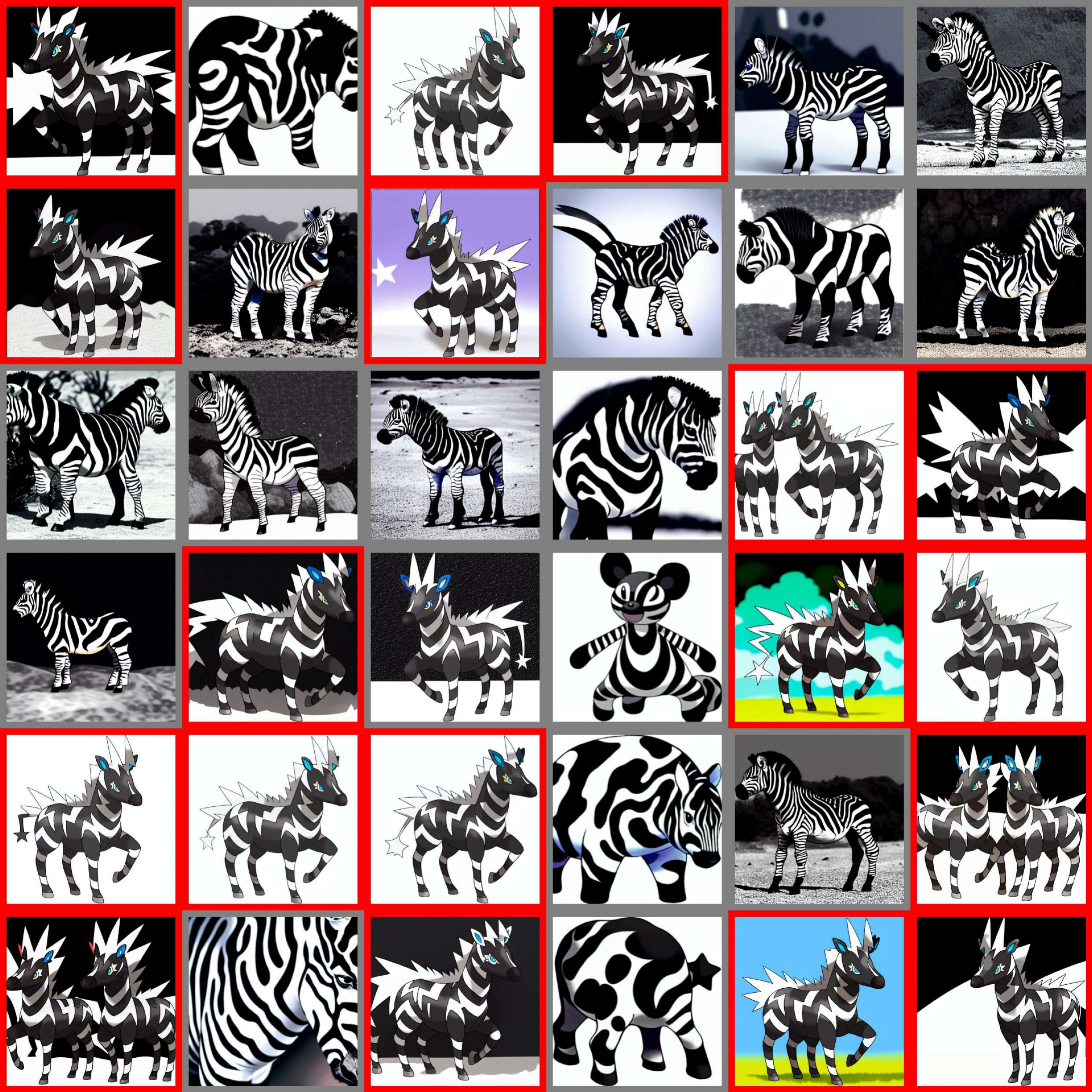}
        \caption{Output w/o CP-$k$.}
    \end{subfigure}
    \hfill
    \begin{subfigure}[t]{0.24\textwidth}
        \includegraphics[width=\textwidth]{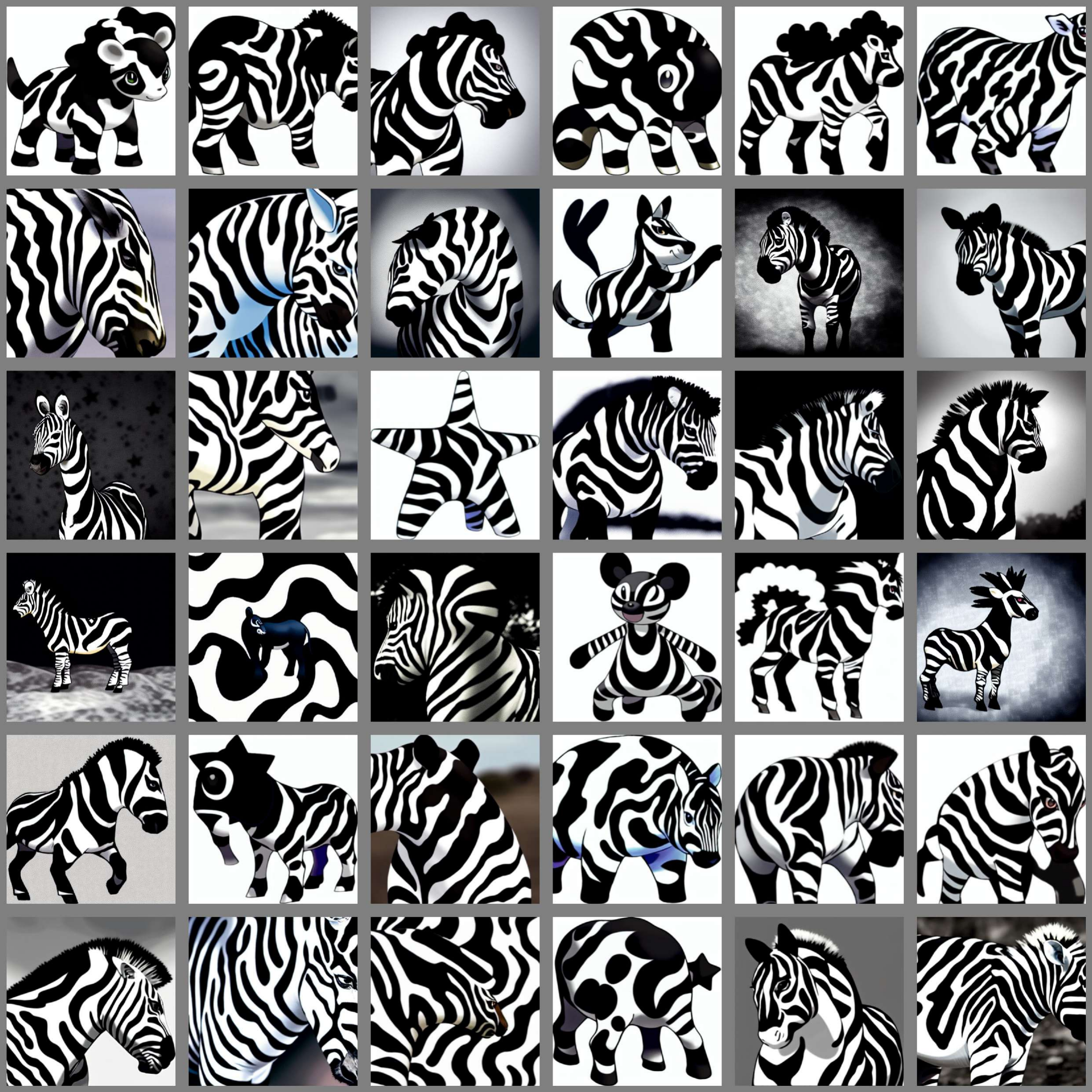}
        \caption{Output w/ CP-$k$.}
    \end{subfigure}
    \hfill
    \begin{subfigure}[t]{0.24\textwidth}
        \includegraphics[width=\textwidth]{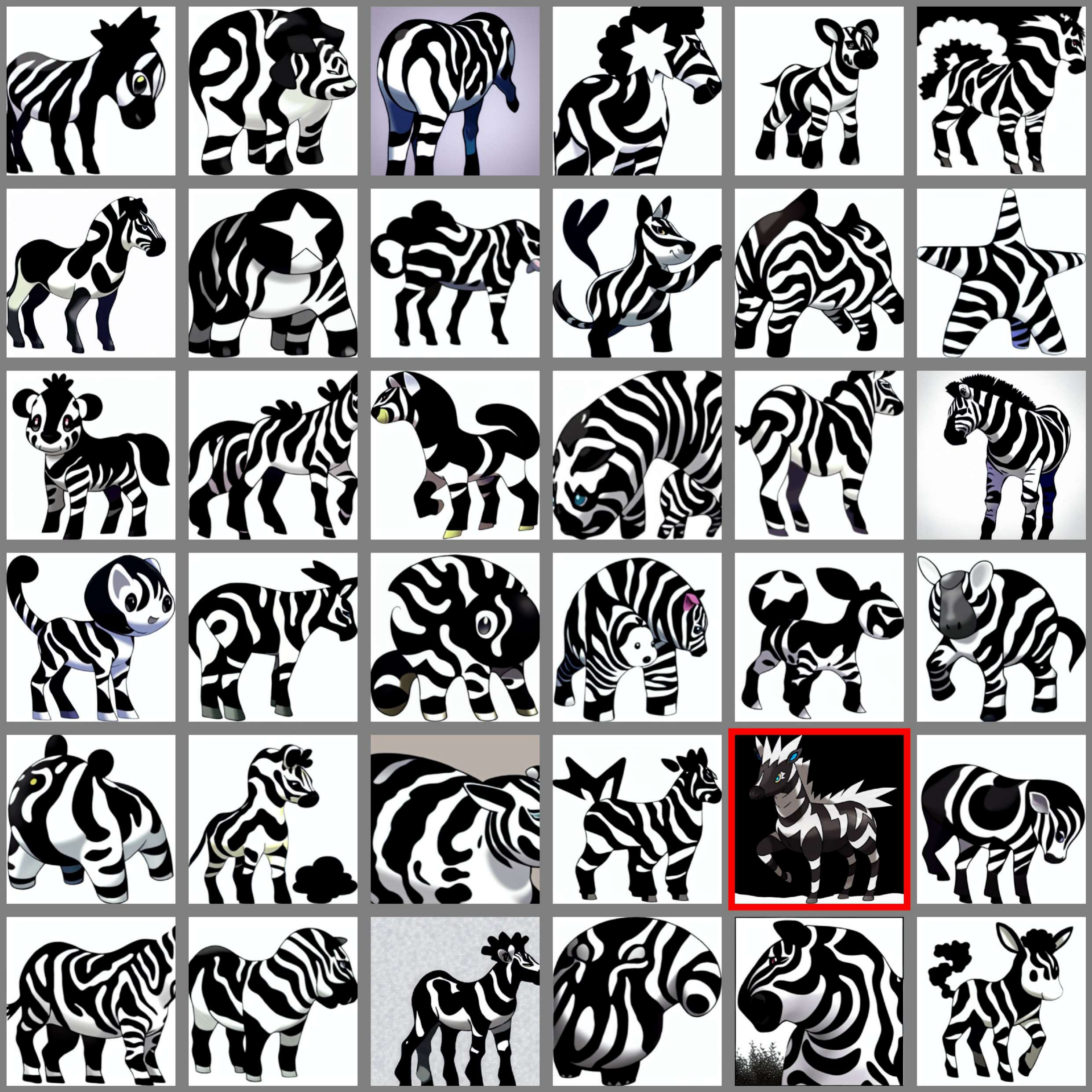}
        \caption{\ourmethodshort output w/ CP-$k$.}
    \end{subfigure}
    \hfill
    \begin{subfigure}[t]{0.24\textwidth}
        \includegraphics[width=\textwidth]{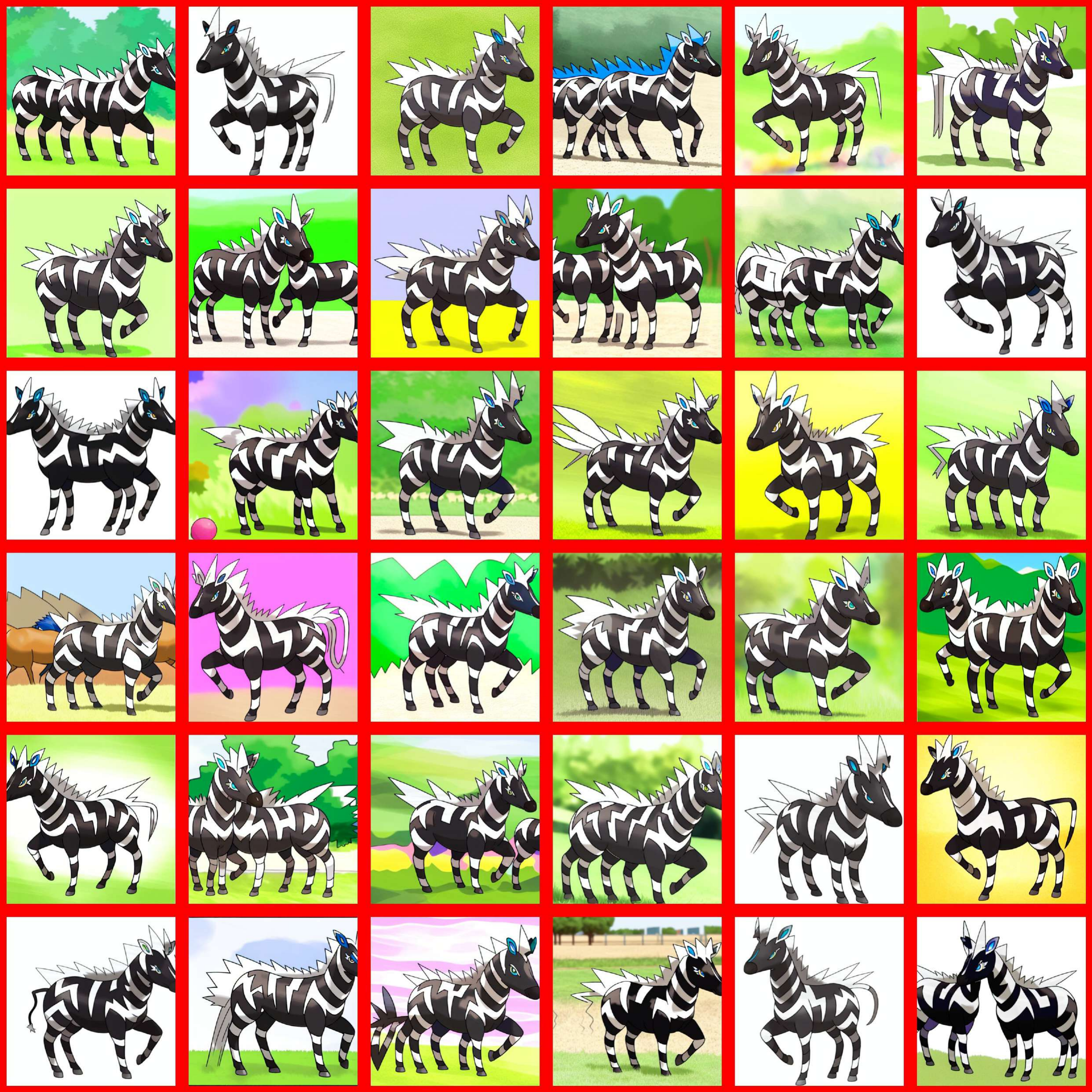}
        \caption{Anti-NAF \ourmethodshort output w/ CP-$k$.}
    \end{subfigure}
    \caption{Example outputs given the copyright image in \cref{fig:demo} as target (potential infringing images are marked with red boundaries).
    In (a), using a benign prompt, we observe a high incidence of infringing content from models without copyright protection (``w/o CP-$k$'').
    In contrast, (b) shows that after applying the copyright protection mechanism (``w/ CP-$k$''), all samples are safe as CP-$k$ rejects all infringing content.
    In (c), we find that amplification (Amp.) attack with a benign prompt results in limited success.
    Notably, by amplification attack with an adversarial prompt obtained from our proposed Anti-NAF algorithm, almost \emph{all} output in (d) are copyright-infringed. }
    \label{fig:example}
\end{figure*}

In recent years, the advancement of large generative models \cite{sohl2015deep,ho2020denoising,song2020score} has revolutionized high-quality image synthesis \cite{rombach2022high,ramesh2022hierarchical,saharia2022photorealistic}, paving the way for commercial applications that enable the public to effortlessly craft their own artworks and designs \cite{gal2023image,saharia2022palette,ruiz2023dreambooth,brooks2023instructpix2pix,kawar2023imagic,lugmayr2022repaint}.
Nevertheless, these models exhibit notable memorization capabilities to produce generations highly similar to the training data \cite{carlini2023extracting}.
This resemblance raises growing concerns about copyright infringement, especially when copyrighted data is used for training \cite{somepalli2023understanding,henderson2023foundation,vyas2023provable,elkin2023can}.

To address these concerns, there has been a surge in research focused on protecting copyrighted data from potential infringement by outputs of generative models \cite{vyas2023provable,liang2023adversarial,shan2023glaze,schramowski2023safe,zhang2023forget,kumari2023ablating,gandikota2023erasing}. 
Among these studies, a pivotal concept involves establishing a probabilistic upper-bound against the generation of infringing content by generative models.
We refer to this suite of approaches as \textbf{probabilistic copyright protection}.
Most notably, Vyas \etal \cite{vyas2023provable} introduce a mathematical definition of copyright known as \textit{near-access freeness} (NAF).
Their method enforces generative diffusion models to exhibit akin behaviors as \emph{safe} models, which has no access to the copyrighted image.
By leveraging the improbability of safe models generating infringing content, the probability of generative models doing the same is thereby substantially reduced.
The copyright protection algorithm of NAF, CP-$k$, can filter out infringing content generated by the models with high probability, even when the input prompts are adversarially designed.

In this paper, we propose Virtually Assured Amplification Attack (VA3), a novel online attack framework, to show the vulnerability of probabilistic copyright protection. 
This framework induces text-to-image generative models with probabilistic copyright protection to generate infringing content. Our approach is grounded in the realization that in real-world scenarios, a malicious attacker intending to induce copyright infringement could engage in multiple interactions with the generative model via prompts.
This persistent engagement poses a significant challenge to probabilistic protection methods, as it \textbf{amplified} the probability of producing infringing content of each generation.

In our proposed framework, the attacker functions as a conditional prompt generator, creating adversarial prompts iteratively based on previous interactions with the generative model. 
Our primary theoretical result, Theorem~\ref{theorem1}, suggests that the amplification attack is guaranteed to succeed with high probability, given a sufficient number of interactions with the generative model and a strictly positive lower-bound on success probability of each single engagement.
Regarding practical algorithms, our work encompasses two technical innovations.
Firstly, we present effective strategies to manage the exploration–exploitation dilemma in online prompt selection, thereby enhancing the stability of the attack.
Secondly, we propose Anti-NAF, a theoretically motivated adversarial prompt optimization algorithm tailored for NAF copyright protection, to generate prompts fulfilling the conditions of Theorem~\ref{theorem1}. 

Our experimental results validate the efficacy of our proposed online attack approach under diverse scenarios.
These findings underline the potential copyright infringement risk of applying probabilistic protection in practical applications of text-to-image generative models, for both providers and users. 

%% file: sec/2_related_work.tex
\section{Related Work}
\label{sec:related_work}
\subsection{Copyright Issues in Generative Models}
Text-to-image generative models, trained on large-scale datasets like LAION \cite{schuhmann2022laion}, have been equipped with enhanced memorization ability to generate outputs of high semantic similarity to their training data \cite{carlini2023extracting, somepalli2023diffusion}.
Given the prevalence of copyrighted works in these datasets, the significant risk of copyright infringement for these generations has raised great concerns from the public \cite{saveri2023stable,chess2022some} and researchers \cite{somepalli2023understanding,vyas2023provable,henderson2023foundation,aboutalebi2023deepfakeart,casper2023measuring,somepalli2023diffusion,scheffler2022formalizing}.
Many efforts have been made to safeguard copyrighted materials from being infringed by generative diffusion models.
Some researchers \cite{liang2023adversarial,liang2023mist,salman2023raising,shan2023glaze} introduced data perturbation, where input data is modified to hinder the model to imitate copyrighted features.
Another separate line of works \cite{schramowski2023safe,kumari2023ablating,gandikota2023erasing,zhang2023forget,liu2023cones} exploited concept removal that erases unsafe concepts from existing pre-trained diffusion models to mitigate the risk of undesirable generations. 
In an alternative approach, researchers studied watermark protection for copyrighted data  \cite{zhao2023recipe,peng2023protecting,wen2023tree,ye2023duaw,cui2023diffusionshield,ray2020recent} to encode ownership information into potentially infringed outputs. 

A notable contribution, Vyas \etal~\cite{vyas2023provable} first provided a mathematical probabilistic upper-bound against copyright-infringed generation.
They asserted that the proposed \emph{near access-freeness} (NAF) offers robust guarantees for copyright protection.
However, Elkin-Koren \etal~\cite{elkin2023can} argued the limitation of this method for reducing copyright to a matter of privacy from a legal perspective.
In this paper, we build upon these discussions to present a significant challenge to these probabilistic copyright protection methods through the amplification attack.


\subsection{Vulnerability of Diffusion Models}
A rising number of works focus on the vulnerability of diffusion models to various types of attacks.
Poisoning attack \cite{chou2023backdoor,chen2023trojdiff,zhai2023text,wu2023proactive,shan2023prompt} studies the problem of manipulating training data to induce unsafe behaviors in diffusion models during the training phase.
In the context of the real-world inference phase for existing text-to-image diffusion models, other works also investigate the robustness and safety against different prompt inputs.
Researchers \cite{zhuang2023pilot, maus2023adversarial, gao2023evaluating, liu2023intriguing} have shown that injecting a slight perturbation to the input prompts will mislead the unprotected model to generate semantically unrelated images.
Furthermore, with carefully crafted prompts, unsafe images can easily evade detection-based safety filters in text-to-image diffusion models \cite{rando2022red,qu2023unsafe}.
Other studies \cite{chin2023prompting4debugging, tsai2023ring} red-teaming concept removal copyright protection methods by finding problematic prompts that can recover erased unsafe concepts to yield undesirable generations.
Diverging from these existing attack approaches on heuristic protection methods, in this paper, our focus is on challenging the vulnerability of probabilistic copyright protection methods.


%% file: sec/3_preliminary.tex
\section{Preliminaries}
\label{sec:preliminaries}

\subsection{Text-to-Image Diffusion Models}

Take DDPM \citep{ho2020denoising} for example, a typical diffusion process consists of a predefined forward process and a reverse process.
Specifically, the forward process corrupts the original data $\mathbf{x}_0\sim q(\mathbf{x})$ into a standard Gaussian noise $\mathbf{x}_T\sim\mathcal{N}(0,\mathbf{I})$ through $T$ timesteps.
In the reverse process, the denoising network $\bm{\epsilon}_\theta$ is learned to denoise the corrupted $\mathbf{x}_t$ by predicting the sampled noise $\bm{\epsilon}\sim \mathcal{N}(0,\mathbf{I})$ that added to $\mathbf{x}_0$.
The objective of the diffusion model can be simplified into the following form:
\begin{equation}
    \mathcal{L}_{\rm{DDPM}}=\mathbb{E}_{t,\mathbf{x}_0,\bm{\epsilon}} \left[ \|\bm{\epsilon}-\bm{\epsilon}_{\theta}(\mathbf{x}_t,t)\|^2\right]
    \label{eq:ddpm}
\end{equation}
Text-to-image diffusion models further use prompts to guide the sampling process for generating desired images.
Take Stable Diffusion \cite{rombach2022high} for example, they incorporate a pre-trained CLIP \cite{radford2021learning} as text encoder $\bm{\tau}$ to encode an input text $\mathbf{y}$, where $\mathbf{c}=\bm{\tau}(\mathbf{y})$.
An image encoder $\mathcal{E}$ is employed to map an input image $\mathbf{x}_0$ into its latent representation $\mathbf{z}_0=\mathcal{E}(\mathbf{x}_0)$.
The training objective is formulated as:
\begin{equation}
    \mathcal{L}_{\rm SD}=\mathbb{E}_{t,\mathbf{z}_0,\bm{\epsilon},\mathbf{c}} \left[ \|\bm{\epsilon}-\bm{\epsilon}_{\theta}(\mathbf{z}_t,t,\mathbf{c})\|^2_2\right]
    \label{eq:sd}
\end{equation}


\subsection{Near Access-Freeness Copyright Protection}

As a pioneering work of probabilistic copyright protection method, Vyas \etal~\cite{vyas2023provable} formally defines Near Access-Freeness \emph{(NAF)} to provide a probabilistic guarantee against the generation of infringing content.
They provide practical algorithms (CP-$k$) to protect an arbitrary generative model $p$ from copyright infringement.
Suppose we have a cover of \emph{safe} models $\mathcal{S}$ satisfying that for any piece of copyrighted data $y_C$, there exists some $q=\emph{safe}_C \in \mathcal{S}$ trained without access to $y_C$.
The copyright protection at a pre-given threshold $k$ can be achieved by sampling $y\sim p(\cdot|x)$ with a prompt $x$ and accepting $y$ if:
\begin{equation}
    \rho(y|x):= \max_{q\in\mathcal{S}} \log \frac{p(y|x) }{q(y|x)}\leq k
    \label{eq:cpfree}
\end{equation}
Intuitively, $\rho(y|x)$ serves as a criterion to distinguish non-infringing content from infringing content, while $k$ determines the threshold for judgment.
Let $p_k$ be the protected model with threshold $k$ and $\mathcal{Y}_C$ be the set of infringing contents, CP-$k$ provides the following probabilistic upper-bound for copyright infringement:
\begin{equation}\label{eq:cp-k_upper_bound}
    p_k(y\in\mathcal{V}_C|x) \le \frac{2^k}{\nu_k(x)}\cdot \emph{safe}_C(y\in\mathcal{Y}_C | x)
\end{equation}
where $\nu_k(x)$ denotes the acceptance rate of $p_k$.
For different prompts $x$, the optimal choice of $k$ changes with varying distribution of $\rho(y|x)$.
In the subsequent sections of our paper, we use $k_x$ to emphasize the dependency.

%% file: sec/4_method.tex
\section{Method}
\label{sec:method}

\subsection{Problem Formulation}\label{sec:prob_form}

In this paper, we consider a text-to-image generative model $\tilde{p}$ equipped with probabilistic copyright protection.
The probability that $\tilde{p}$ generates infringing content with prompt $x$ is upper-bounded.
Following the typical adversarial prompt attack setting, we consider a malicious attacker who seeks to manipulate $\tilde{p}$ to produce content that violates the copyright of a specific piece of target copyrighted data $y_C$, of which the infringing contents form $\mathcal{Y}_C \subseteq \mathcal{Y}$.

Different from the standard paradigm, we consider an online attack scenario where the attacker is allowed to interact with $\tilde{p}$ for $T$ times within each attack trial.
During the $t=1,\cdots,T$-th interaction, the attacker inputs a prompt $x_t$ and receives a generated sample $y_t \sim \tilde{p}(\cdot|x_t)$.
Besides furnishing a prompt, the attacker is prohibited from intervening directly in the generation process.
This constraint is consistent with real-world scenarios where users interact with generative models as black boxes through APIs.
At the end of each attack trial, the attacker will select $y^* \in \{y_t\}_{t=1}^T$ as the final output, and the attack is regarded successful if $y^* \in \mathcal{Y}_C$.
The formal objective of the attacker is to maximize the success rate of attack, \ie,

\begin{equation}\label{eq:primal_obj}
    \max_{x\in\mathcal{X}} \ P_{y\sim\tilde{p}(\cdot|x)}(y^* \in\mathcal{Y}_C)
\end{equation}

\subsection{Virtually Assured  \Ourmethod Attack on Probabilistic Copyright Protection}
\label{sec:amplification}

Copyright protection mechanisms in generative models, \eg, \emph{NAF}, are designed to yield a low upper-bound on the probability of generating infringing samples.
Such protection is confined to single or infrequent generations. 
However, in scenarios where malicious attackers engage in multiple targeted generation requests, the probability of a successful attack can be \textbf{amplified}.

We propose a universal online \ourmethod attack framework, detailed in \cref{alg0}, for a virtually assured success against probabilistic copyright protection.
In this framework, the attacker is modeled as an adversarial prompt generator $\mathcal{A}(x_t|\left<(x_s, y_s)_{s=1}^{t-1}\right>)$, which iteratively generates the prompt $x_t$ for the current step $t$ based on the interaction history with the generative model.
For selecting the optimal sample, we employ a scoring function $\mathcal{S}: \mathcal{Y} \rightarrow \mathbb{R}$ to evaluate all the samples and return the highest-scoring sample as the final result.
An attack is deemed successful if the score of the returned result is higher than the target score $\mathcal{S}_{tar}$.
For example, an essential choice of the scoring function could be the indicator function $\mathbb{I}(y \in \mathcal{Y}_C)$, with the target score $S_{tar}=0$, which ensures that any infringing sample generated will be returned as the final result, making the attack as successful.
In practice, to reduce the manual effort in identifying copyright infringement, a computable surrogate scoring function with a target score may be utilized to establish a standard for copyright infringement.

We introduce the following theorem, which suggests the virtually assured success of the \ourmethod attack with an assumption on the lower-bound of success probability for each single-shot attack attempt.

\begin{theorem}
\label{theorem1}
    Following the notations in \cref{alg0}, for any $\varepsilon \in (0,1)$, the attack is successful with probability at least $1 - \varepsilon$ if $T > \log_{1-\sigma} \varepsilon$, where $\sigma > 0$ is a strictly positive lower-bound on the success probability shared by every single attack.
\end{theorem}

The intuition behind the \ourmethod attack and \cref{theorem1} is straightforward. 
If the attacker is granted sufficient sampling opportunities with a set of prompts that yield even a modest lower-bound on the single success probability, as the number of sampling attempts increases, the probability of generating infringing samples accumulates at an exponentially fast rate, eventually making an attack almost guaranteed to succeed.
For instance, suppose our attack possesses a relatively low single-shot success probability, say 1\%. 
Provided that we are permitted to repeat the attack approximately $\log_{0.99} 0.01 \approx 459$ times, the probability of at least one successful instance is then amplified to 99\%.

\begin{algorithm}[t]
\caption{\Ourmethod Attack on Probabilistic Copyright Protection}\label{alg0}
\begin{algorithmic}[1]
    \Require Generative model $\tilde{p}$ with probabilistic copyright protection, target copyrighted data $\mathcal{C}$, adversarial prompt generator $\mathcal{A}$, maximum number of steps $T$, score function $\mathcal{S}$.
    \For{$t=1,\dots,T$}
        \State Sample prompt $x_t \sim \mathcal{A}(\cdot|\left<(x_s, y_s)\right>_{s=1}^{t-1})$
        \State Feed $x_t$ to $\tilde{p}$ and receive $y_t \sim \tilde{p}(\cdot|x_t)$
    \EndFor \\
    \Return $y^* = \arg\max_{y \in \{y_{1:T}\}} \mathcal{S}(y)$
\end{algorithmic}
\end{algorithm}

\subsection{Online Prompt Selection}
\label{sec:bandit}
In this section, we consider a specific scenario within the framework described in \cref{sec:amplification}, where the conditional prompt generator (the attacker) is restricted to select a prompt among $K$ candidate prompts $\{x^1,\cdots,x^K\}$, based on previous choices and the scores of received samples.
The generation of candidate prompts can either be independent of or tailored specifically for certain models and copyright protection mechanisms, as we will discuss in \cref{sec:anti-naf}, designed for diffusion model and NAF copyright protection.

Let $a_t \in \{1,\cdots,K\}$ denote the decision at step $t$ and $\pi$ denote the policy of prompt selection, the prompt generation procedure can be formally described as
\begin{equation}
    x_t = x^{a_t}, a_t \sim \pi(\cdot|\left<(x_s, r_s=\mathcal{S}(y_s))\right>_{s=1}^{t-1}).
\end{equation}
In this context, online prompt generation is reformulated as a variant of the multi-armed bandit problem, where each candidate prompt is conceptualized as an arm, and the score of the corresponding sample is the reward for pulling the arm.
Unlike the classic multi-armed bandit framework, our goal is to maximize the probability of obtaining a reward exceeding a specified threshold $S_{tar}$ at least once within $T$ trials.
This involves the trade-off between exploration and exploitation.
Herein, we present two variants of the $\varepsilon$-greedy algorithm.
Initially, akin to the conventional algorithm, we conduct $m \le \lfloor T / K \rfloor$ trials for each arm.
For the rest $T - mK$ steps, a random action will be taken with probability $\varepsilon > 0$ for exploration.
Otherwise, the best action $\hat{a}_t^* = \argmax_a \hat{Q}_t(a)$ according to the evaluation $\hat{Q}_t(a)$ will be taken for exploitation.
The choice of $\hat{Q}_t$ diverges into two distinct variants.

\noindent \textbf{$\varepsilon$-greedy-max} \quad Inspired by the fact that only the max reward matters, the maximum reward received so far is employed for evaluation, \ie, $\hat{Q}_t(a) = \max\{r_s: a_s=a\}_{s=1}^t$.

\noindent \textbf{$\varepsilon$-greedy-cdf} \quad Essentially, we aim to maximize the probability that the reward for the next step exceeds a threshold $S_{tar}$.
If we assume that the reward distribution for each arm adheres to a normal distribution, we have $\hat{Q}_t(a)=1 - \Phi(\frac{S_{tar} - \hat{\mu}_t(a)}{\hat{\sigma}_t(a)})$, where $\hat{\mu}_t(a)=\frac{1}{N_t(a)}\sum_{s=1}^t r_s\mathbb{I}(a_s=a)$, $\hat{\sigma}_t^2(a)=\frac{1}{N_t(a)-1}\sum_{s=1}^t (r_s\mathbb{I}(a_s=a) - \hat{\mu}_t(a))^2$, and $\Phi$ denotes the cumulative distribution function (cdf) of standard normal distribution.

Our empirical findings suggest that a well-calibrated balance between exploration and exploitation can enhance the stability of the attack.

\subsection{Anti-NAF: Adversarial Prompt Optimization Against NAF Copyright Protection}
\label{sec:anti-naf}

In this section, we narrow the scope of discussion to adversarial prompt discovery against CP-$k$, a \emph{NAF}-based copyright protection algorithm.
CP-$k$ modifies the probability density function of an unprotected generative model $p$ as
\begin{equation}\label{eq:aug_gen_model}
\tilde{p}(y|x) \propto p(y|x) \mathbb{I}(\rho(y|x) \le k_x)
\end{equation}
where $\rho$ is defined in \cref{eq:cpfree} and $k_x$ is a prompt-dependent threshold determined by the generative system, which is typically inaccessible to the user.
CP-$k$ provides a probabilistic upper-bound for copyright infringement in \cref{eq:cp-k_upper_bound}.

Building upon our initial assumptions, we further assume that the attacker is granted full access, \ie., knowing the structures and parameters of these models, to the unprotected generative model $p$ and safe models $q\in\mathcal{S}$.
While this white-box setting renders our attack narrowed to open-source models and those at risk of backend leakage, the potential threat of our attack remains substantial, as it does not require intervention in the generation process.
This means that if a malicious attacker gains access to a leaked model and discovers adversarial prompts, any other users, lacking access to the underlying models, can readily replicate these adversarial prompts to reproduce the attack. 

Recall that our objective is to find adversarial prompts capable of inducing models protected by CP-$k$ to generate infringing content with a strictly positive probability, \ie, 
\begin{equation}\label{eq:far}
    \max_{x\in\mathcal{X}} P_{y\sim \tilde{p}(\cdot|x)}(y\in\mathcal{Y}_C)
\end{equation}
We deduce a lower-bound for the infringement probability with optimal adversarial prompt in \cref{theorem2}, subsequently guiding the development of a practical algorithm.

\begin{definition}[Local Continuity]\label{def1}
    Given a distance measure $\mathcal{D}$ defined in $\mathcal{Y}$, a model $p$ is called $(\epsilon, \alpha)$-local-continuous around $y_0\in\mathcal{Y}$ if for any prompt $x\in\mathcal{X}$, there exists $\epsilon, \alpha > 0$ such that for any $y\in\mathcal{B}_\mathcal{D}(y_0, \epsilon):= \{y \in \mathcal{Y}: \mathcal{D}(y_0, y) < \epsilon\}$, $|p(y_0|x) - p(y|x)| < \alpha \mathcal{D}(y_0, y)$.
\end{definition}

\begin{theorem}\label{theorem2}
    Assume there is a distance measure $\mathcal{D}$ defined in $\mathcal{Y}$ such that (i) $p$ is $(\epsilon_p, \alpha)$-local-continuous around $y_C$, (ii) every $q \in \mathcal{S}$ is local-continuous around $y_C$, and (iii) there exists $\epsilon_c > 0$ such that $\mathcal{B}_\mathcal{D}(y_C, \epsilon_c) \subseteq \mathcal{Y}_C$. The objective defined in \cref{eq:far} has the following lower-bound for any $\eta, \delta > 0$,
    \begin{equation}
    \max_{x\in\mathcal{X}} P_{y\sim \tilde{p}(\cdot|x)}(y\in\mathcal{Y}_C)\\
    \ge \max_{x\in\tilde{\mathcal{X}}_{\eta, \delta}} \ \eta C_1 - \alpha C_2
    \end{equation}
    where $\tilde{\mathcal{X}}_{\eta, \delta} = \{x\in\mathcal{X}: p(y_C|x) \ge \eta, \rho(y_C|x) < k_x - \delta\}$ and $C_1, C_2$ are constants independent on $x$ given as
    \begin{equation}
        \begin{aligned}
            C_1 = \int_{y\in \mathcal{B}_{\mathcal{D}}(y_C, \epsilon)} dy, \ \ C_2 = \int_{y\in \mathcal{B}_{\mathcal{D}}(y_C, \epsilon)} \mathcal{D}(y_C, y)dy,
        \end{aligned}
        \nonumber
    \end{equation}
    where $\epsilon = \min(\epsilon_p, \epsilon_c, \epsilon_\rho)$ with $\epsilon_\rho:={\rm inf}_{x\in\tilde{\mathcal{X}}_{\eta, \delta}}{\rm sup}\{\epsilon: \rho(y|x) < k_x, \forall y \in \mathcal{B}_{\mathcal{D}}(y_C, \epsilon)\}.$
\end{theorem}

To ensure that the lower-bound in the Theorem~\ref{theorem2} is nontrivial, we need to set $\eta > \alpha C_2 / C_1$ and search along prompts satisfying (a) $\rho(y_C|x) < k_x$, (b) $p(y_C|x) \ge \eta$.
Unfortunately, the mechanism to determine the threshold $k_x$ is assumed inaccessible to the attacker. 
To obtain a feasible optimization objective, we instead minimize $\rho(y_C|x)$ subject to $p(y_C|x) \ge \eta$, as an alternative.
The optimization objective then becomes

\begin{equation}
         \min_x \rho(y_C|x) \quad {\rm s.t.} \ p(y_C|x)\geq \eta.
    \label{eq:obj}
\end{equation}

\begin{algorithm}[t]
\caption{Anti-NAF: Adversarial Prompt Optimization Against NAF Copyright Protection}\label{alg1}
\begin{algorithmic}[1]
\Require Denoising network $\bm{\epsilon}_{\theta_p}, \bm{\epsilon}_{\theta_{q}}$, text encoder ${\bm{\tau}}$, target $y_C$, denoising steps $T$, optimization step $S$, loss clip bound $\varphi$, loss weight $\lambda$, learning rate $\gamma$
    \State $\mathbf{P}=\{\mathbf{e}_1, \dots, \mathbf{e}_n\}\sim \bm{E}^{|\mathcal{V}|\times d}$
    \For{$1,\dots, S$}
        \State $x={\rm Proj}(\mathbf{P})$
        \State $\triangleright$ Diffusion reconstruction task
        \State $t\sim{\rm Uniform}(\{1,\dots,T\})$
        \State $\bm{\epsilon}\sim \mathcal{N}(\bm{0}, \mathbf{I})$ 
        \State $y_t=\sqrt{\bar{\alpha}_t}y_C+\sqrt{1-\bar{\alpha}_t}\bm{\epsilon}$ 
        \State $\mathcal{L}_p = \|\bm{\epsilon}-\bm{\epsilon}_{\theta_p}(y_t,t,\bm{\tau}(x))\|^2_2$
        \State $\mathcal{L}_{q} = \|\bm{\epsilon}-\bm{\epsilon}_{\bm{\theta}_{q}}(y_t,t,\bm{\tau}(x))\|^2_2$\  for all $q\in\mathcal{S}$
        \State $\triangleright$ Calculate the gradient w.r.t projected embedding
        \State $g=\nabla_x(\lambda\cdot\max(\mathcal{L}_p, \varphi) + (1-\lambda)\cdot\max_{q\in\mathcal{S}}(\mathcal{L}_{q}))$
    \State $\triangleright$ Apply the gradient on continuous embedding 
        \State $\mathbf{P}=\mathbf{P}-\gamma g$ 
    \EndFor \\
    \Return ${\rm Proj}(\mathbf{P})$
\end{algorithmic}
\end{algorithm}

We follow the reconstruction task of the original text-to-image diffusion models for direct optimization. 
In aligning with the attack objective of \cref{eq:obj}, we develop the optimization objective from two aspects: 
\begin{equation}
    \mathcal{L}_p = \mathbb{E}_{t,\bm{\epsilon},x}[\|\bm{\epsilon}-\bm{\epsilon}_{\theta_p}(y_t,t,\bm{\tau}(x))\|^2_2]
\end{equation}
\begin{equation}
    \mathcal{L}_{q} = \mathbb{E}_{t,\bm{\epsilon},x}[\|\bm{\epsilon}-\bm{\epsilon}_{\theta_{q}}(y_t,t,\bm{\tau}(x))\|^2_2] 
\end{equation}
Where $y_t$ is the noisy version of target copyrighted data $y_C$ at denoising step $t$; $\bm{\epsilon}_{\theta_p}, \bm{\epsilon}_{\theta_{q}}$ is the denoising network of generative model $p$ and safe model $q\in\mathcal{S}$ respectively. 

While $\mathcal{L}_p$ is designed to increase the possibility of model $p$ in generating the desired infringed content $y_C$, the minimization of $\mathcal{L}_{q}$ corresponds to the maximization of $q(y_C|x)$, consequently leading to the reduction of $\rho(y_C|x)$. 
Hence, the overall optimization objective can be formulated as a weighted sum of $\mathcal{L}_{p}$ and $\mathcal{L}_q$:


\begin{equation}
    \mathcal{L} = \lambda\cdot\max(\mathcal{L}_p, \varphi) + (1-\lambda)\cdot\max\limits_{q\in\mathcal{S}}(\mathcal{L}_{q})
    \label{eq:loss}
\end{equation}
Where $\varphi$ is a loss clip to mitigate the conflict between $\mathcal{L}_{p}$ and $\mathcal{L}_q$ in minimizing $\rho$.


The complete Anti-NAF prompt optimization process is detailed in \cref{alg1}.
We conduct optimization in the continuous embedding space with a sequence of learnable embeddings $\mathbf{P}=\{\mathbf{e}_1, \dots, \mathbf{e}_n|\mathbf{e}_i\in\mathbb{R}^d\}$, where $n$ is the sequence length and $d$ is the embedding dimension.
Given that each word token $w$ in a vocabulary $\mathcal{V}$ can be represented as a corresponding embedding $\textsc{Emb}(w)$ using an embedding matrix $\bm{E}^{|\mathcal{V}|\times d}$, each embedding $\mathbf{e}$ in the sequence can be mapped to its nearest token embedding in $\mathcal{V}$ under some similarity metric, \eg, cosine similarity.
The projection function can be defined as ${\rm Proj}(\mathbf{e})=\argmin_{\substack{w\in \mathcal{V}}} \cos(\textsc{Emb}(w), \mathbf{e})$.
The resultant text prompt is a sequence of these projected tokens: $x={\rm Proj}(\mathbf{P})=\{{\rm Proj}(\mathbf{e}_1), \dots, {\rm Proj}(\mathbf{e}_n)\}$.
In addition, following the approach in \cite{wen2023hard}, the continuous embeddings $\mathbf{P}$ are projected into discrete tokens $x$ for each forward pass in every optimization step.
The gradient of the projected prompt $x$ is then applied to update the continuous embeddings $\mathbf{P}$.

%% file: sec/5_exp.tex
\begin{figure*}[ht!]
    \centering
    \includegraphics[width=\linewidth]{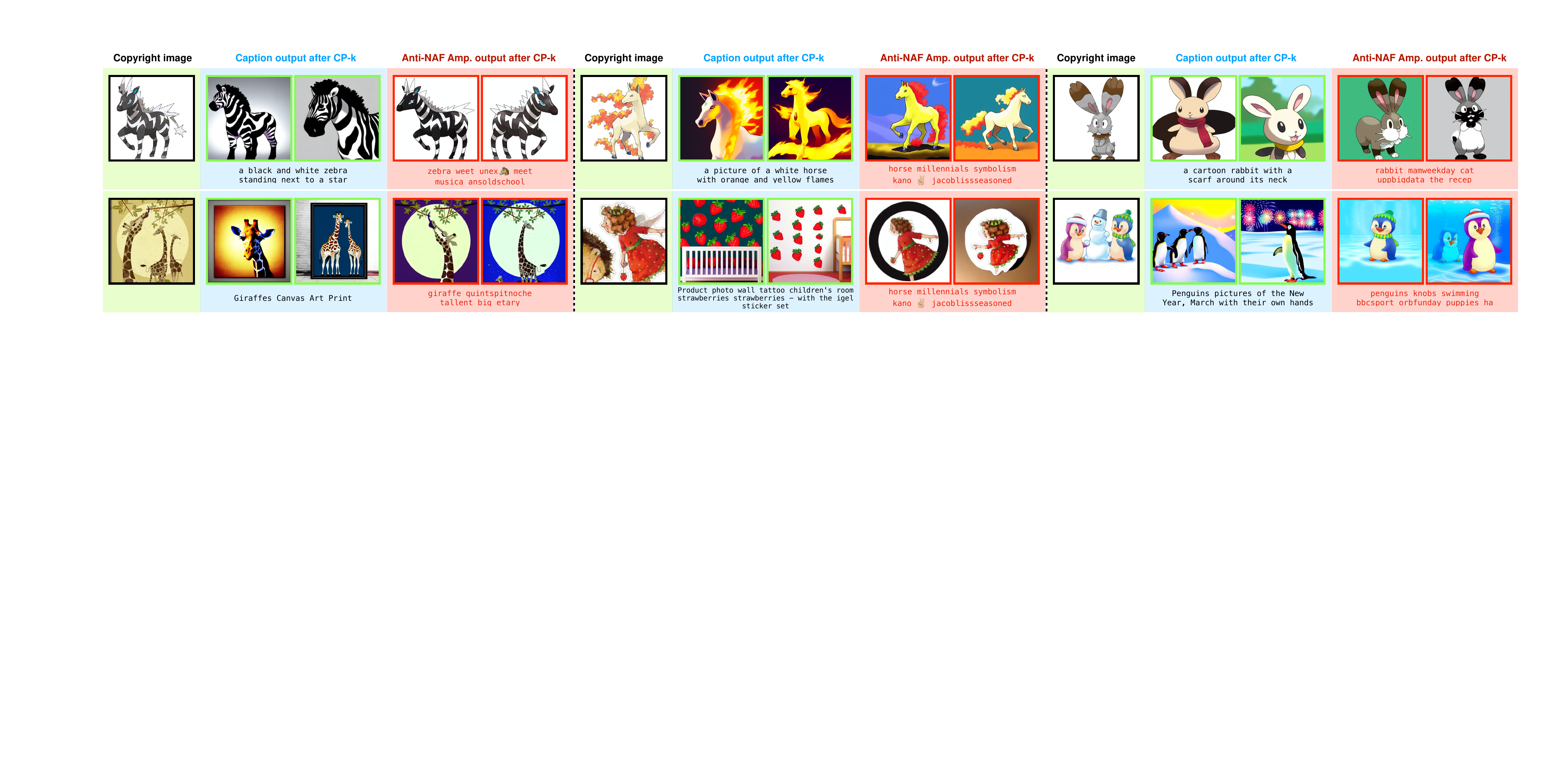}
    \caption{Visualization of generated images on different copyright targets. The examples in the first and second rows are selected from POKEMON and LAION-mi respectively. The prompts used to generate output are given below each group of images. Remarkably, the copyright-infringed content generated with Anti-NAF amplification reveals the vulnerability of probabilistic copyright protection CP-$k$.}
    \label{fig:selected_examples}
\end{figure*}

\section{Experiments}
\label{sec:exp}

\subsection{Experimental Settings}

\subsubsection{Evaluated Datasets and Models}

Since training large text-to-image diffusion models from scratch is impractical, we fine-tune the pre-trained StableDiffusion-v1-4 model provided by Huggingface\footnote{https://huggingface.co/CompVis/stable-diffusion-v1-4} with two datasets: POKEMON \cite{pinkney2022pokemon} and LAION-mi \cite{dubinski2023towards}.
Note that both datasets do \emph{not} overlap with the pre-training dataset of Stable Diffusion, ensuring that the safe models fine-tuned on them are inaccessible to copyrighted data.
Following \cite{vyas2023provable}, each dataset is split into two disjoint shards $\mathcal{D}_1$ and $\mathcal{D}_2$ to train generative model $q_1$ and $q_2$ respectively.
For copyrighted data $y_C$, $q_2$ is served as safe model $\emph{safe}_C$ if $y_C\in \mathcal{D}_1$ but $y_C\notin \mathcal{D}_2$.
For better experimental illustrations, each copyrighted data is repeated to make up 1\% of the dataset shard.

\textbf{POKEMON} \cite{pinkney2022pokemon}. 
This dataset consists of 833 image-caption pairs, where captions are obtained from the BLIP model \cite{li2022blip}. 
We evaluate our attack on 5 copyrighted images added to one of the shards separately and fine-tune models $q_1$ and $q_2$ on each shard for 5000 steps.

\textbf{LAION-mi} \cite{dubinski2023towards}. 
The dataset is originally constructed for membership inference attacks on diffusion models. 
We only use the nonmembers part of this dataset, which holds a similar data distribution but is disjoint with the pre-training dataset of Stable Diffusion. 
We evaluate our attack on 5 selected copyrighted images, which are separately combined with one of non-copyrighted data shards. Each dataset shard is of size 5000. We fine-tune model $q_1$ and $q_2$ for 25,000 steps. Details for fine-tuning can be found in the appendix.

\subsubsection{Implementation details} 
For the sampling setting of text-to-image diffusion models, images of size $512\times512$ are generated using a classifier-free guidance scale of 7.5 and 50 sampling steps with the default scheduler. 
The \ourmethod step is set to 100 and 500 for POKEMON and LAION-mi respectively.
For the prompt optimization process, the learning rate is set to 0.01 using Adagrad \cite{duchi2011adaptive} optimizer with 25,000 optimization steps, the gradient accumulation step is set to 5, the loss clip $\varphi$ for model $p$ is 0.01, and the loss weight $\lambda$ is $0.95$. 
The length of optimized prompts is set to $8$ tokens. All experiments are conducted on A100 GPUs.

\begin{table*}[ht!]\footnotesize
    \centering
    \begin{tabular}{l|c|cc|c|ccc}
    \toprule
    \multirow{2}*{Methods} & \multicolumn{3}{c|}{POKEMON} & \multicolumn{4}{c}{LAION-mi}\\
    \cline{2-8}
    ~ & CIR & FAR@5\%AR$\uparrow$ &FAR@15\%AR$\uparrow$ & CIR &  FAR@10\%AR$\uparrow$ & FAR@30\%AR$\uparrow$ & FAR@50\%AR$\uparrow$ \\
    \midrule
    
    Caption (w/o \ourmethodshort)&	47.00\%& 0.40\%&	3.28\%&	
    49.64\%&  0.00\%&0.00\% &	0.04\% \\
    
    CLIP-Int. (w/o \ourmethodshort)&	26.92\% & 0.84\%&	2.20\% &
    48.12\%&	0.84\%&1.52\% &	2.96\% \\
    
    PEZ (w/o \ourmethodshort)&	7.80\% &1.32\%&	2.76\%&
    15.80\%&0.08\%&0.24\% &	0.40\% \\
    
    Anti-NAF (w/o \ourmethodshort)&	12.88\% &\textbf{ 8.52\%}&	\textbf{10.00\%}&
    33.84\%& \textbf{2.64\%} & \textbf{4.16\%} &	\textbf{7.00\%} \\
    \cline{1-8}
    
    Caption (w/ \ourmethodshort)&	100.00\%& 13.64\%&	
    40.16\%&	100.00\%&	0.00\%&0.00\% &	14.64\% \\
    
    CLIP-Int. (w/ \ourmethodshort)&	99.72\%& 22.84\%&	47.32\%&	
    100.00\%    &	34.84\%   &   37.12\% &	75.60\% \\
    
    PEZ (w/ \ourmethodshort)&	66.92\%& 22.80\%&	37.48\%&	
    98.16\% &	38.04\%    &   50.68\% &	59.64\%	\\
    
    Anti-NAF (w/ \ourmethodshort)&	99.84\%& \textbf{77.36\%}&	\textbf{91.36\%}&
    100.00\%&	\textbf{56.12\%}& \textbf{73.32\%} &	\textbf{95.92\%} \\
    
    
    
   
    \bottomrule
    \end{tabular}
    \caption{Quantitative results. 
    The performance with the amplification attack (``w/ Amp.'') is significantly superior to scenarios without amplification (``w/o Amp.''). Additionally, our proposed Anti-NAF demonstrates notably promising outcomes for providing a substantial probability of copyright infringement when probabilistic protection is applied.
    (CLIP-Int. is the abbreviation for CLIP-Interrogator).}
    \label{tab:main_res}
\end{table*}

\subsection{Evaluation Settings}
We evaluate with a model $p$ behaves with an equal chance of sampling from either model $q_1$ or $q_2$.
We consider such model $p$ as a strict protection model that even with the original caption as prompt, there is a 50\% chance of generating from the safe model. 
In contrast, a model $p$ trained on the entire dataset could frequently produce highly similar generations to the target copyrighted image, making it extremely inefficient for the selection of a valid threshold $k$. 
Next, we will separately discuss the data, the metrics, and the threat prompts for evaluation.

\textbf{Evaluation Data.}  We use samples generated by $p$ for evaluation. 
To determine whether they infringe the copyright of the target image, we rely on the similarity between samples and the target image as there is no widely acknowledged computable standard for infringement to our knowledge.
We employ the current SOTA similarity metric \emph{SSCD} \cite{pizzi2022self} for copy detection.
As SSCD scores of infringed samples for different target images differ significantly by observation (shown in \cref{fig:sim_comp}), we do not indicate a fixed score threshold for all target images.
Instead, we use relative thresholds determined as percentiles of the similarity scores among samples generated by $p$ with the original caption of the target image as the prompt, \eg~SSCD-50\%.
Recognizing the variable criteria of infringement, we report results with other choices of thresholds in \cref{sec:add_results}.

\textbf{Evaluation Metrics.}
The CP-$k$ method achieves copyright protection by selectively accepting generated samples using the threshold $k_x$, which can be indicated by the Acceptance Rate (AR). 
For a given AR, a good protection system is expected to rarely accept infringing content, \ie, have a low False Accept Rate (FAR), as defined in \cref{eq:far}.
It is worth noting that the choice of $k_x$ dictates the trade-off between model safety and efficiency.
Furthermore, to our knowledge, there is no principled way to determine $k_x$.
As a result, we evaluate the success of attack by reporting the FAR at different AR, \eg, FAR@5\%AR.
Additionally, the copyright infringement rate (CIR) is also presented for scenarios \emph{without} copyright protection, \ie, AR=100\%.  

\textbf{Threat Prompts.}
Given that generating images similar to the target image is a necessary condition for a successful attack, we focus on the following threat prompts:
(i) \emph{Target caption}: the original caption of the target image; 
(ii) \emph{PEZ}~\cite{wen2023hard}: a gradient-based discrete optimization approach to discover prompts semantically similar to the target image; 
(iii) \emph{CLIP-Interrogator}~\footnote{https://github.com/pharmapsychotic/clip-interrogator}: a prompt consists of the BLIP caption of the target image and top-$k$ keywords greedily sampled from a keywords collection;
(iv) \emph{Anti-NAF}: an adversarial prompt against NAF copyright protection as described in \cref{sec:anti-naf}.
For each threat prompt, we sample 6400 images from model $p$ for evaluation. 
For the online prompt selection setting, we take the aforementioned three prompts (exclude the target caption) as a prompt candidate set and consider \emph{$\varepsilon$-greedy-max/-cdf Bandit \Ourmethod} as bandit strategies as described in \cref{sec:bandit}.
\begin{figure}[ht!]
    \centering
    \begin{subfigure}[t]{0.49\linewidth}
        \includegraphics[width=\linewidth]{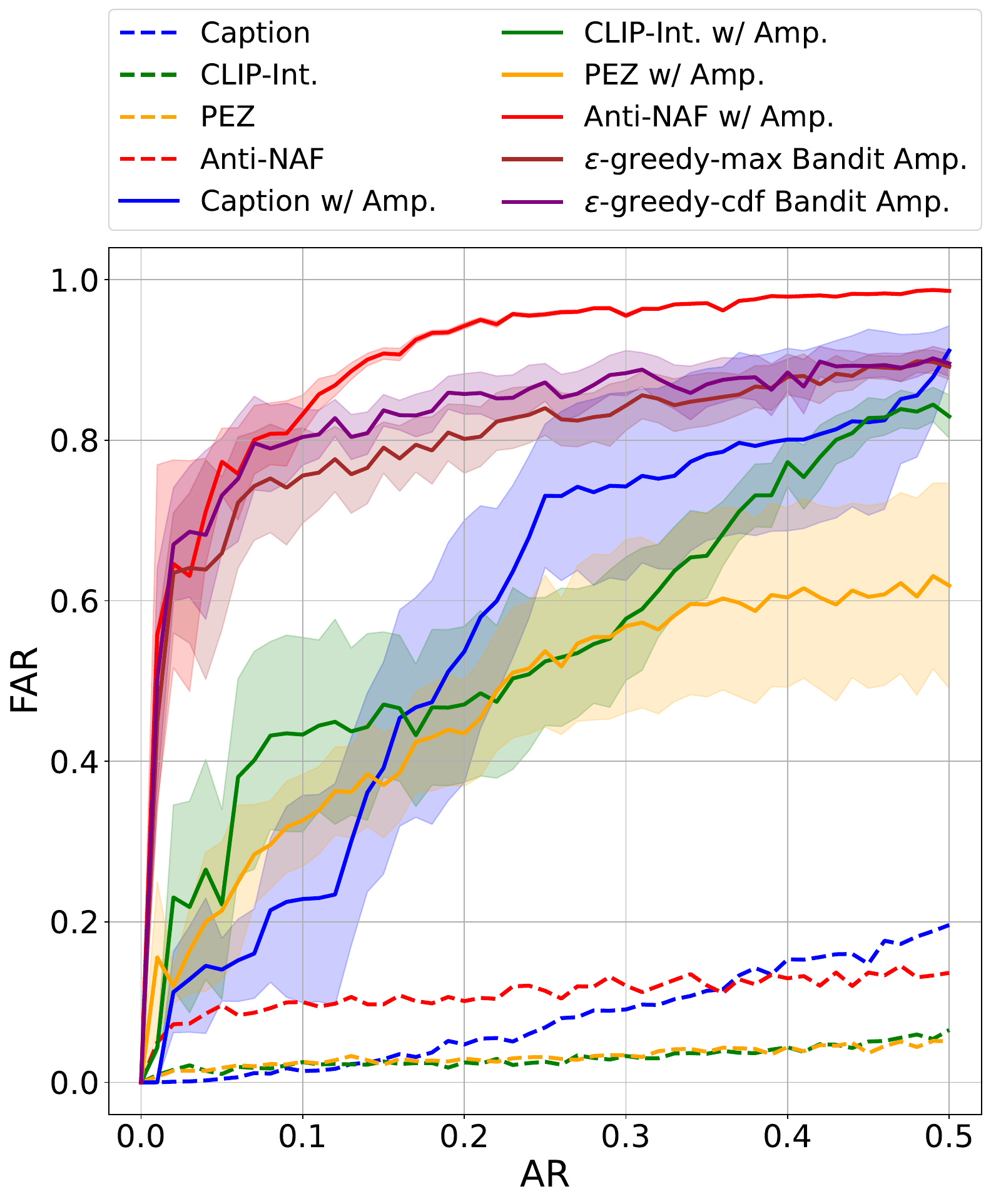}
        \caption{FAR-AR on POKEMON.}
        \label{fig:pokemon_plot}
    \end{subfigure}
    \hfill
    \begin{subfigure}[t]{0.49\linewidth}
        \includegraphics[width=\linewidth]{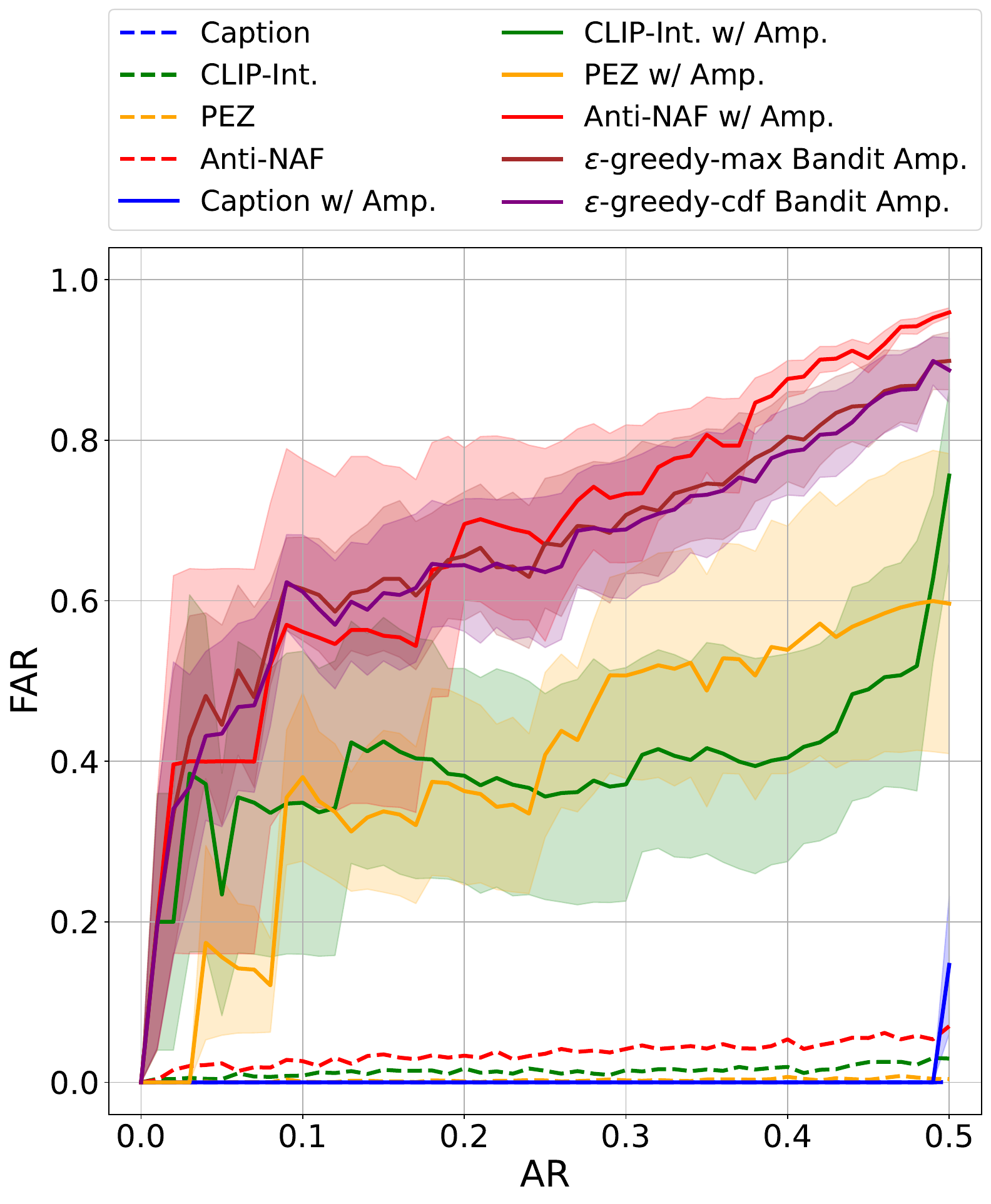}
        \caption{FAR-AR on LAION-mi.}
        \label{fig:laion_plot}
    \end{subfigure}
    \hfill
    \begin{subfigure}[t]{0.49\linewidth}
        \includegraphics[width=\linewidth]{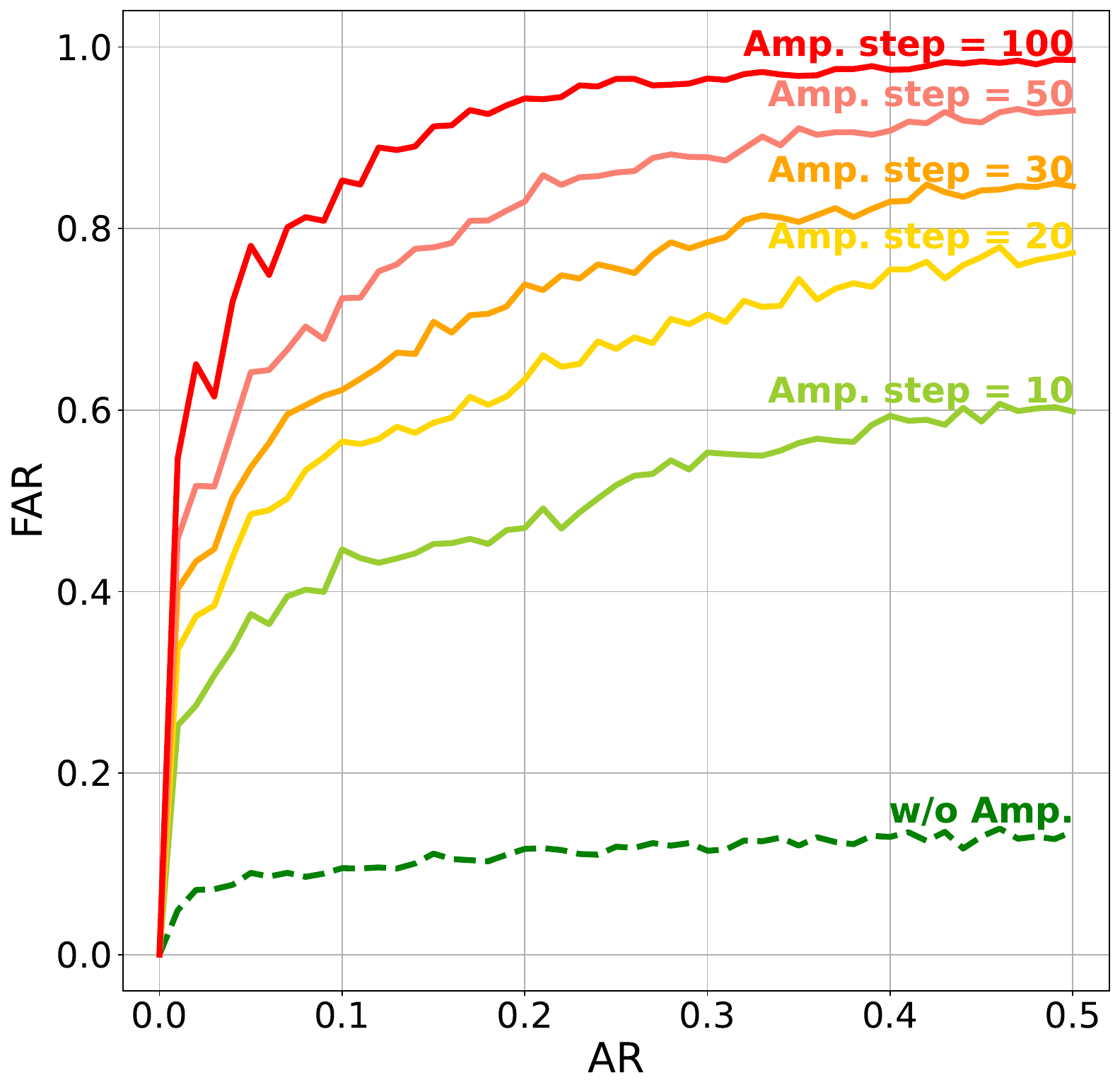}
        \caption{FAR-AR of Anti-NAF with different Amp. steps on POKEMON.}
        \label{fig:pokemon_amp}
    \end{subfigure}
    \hfill
    \begin{subfigure}[t]{0.49\linewidth}
        \includegraphics[width=\linewidth]{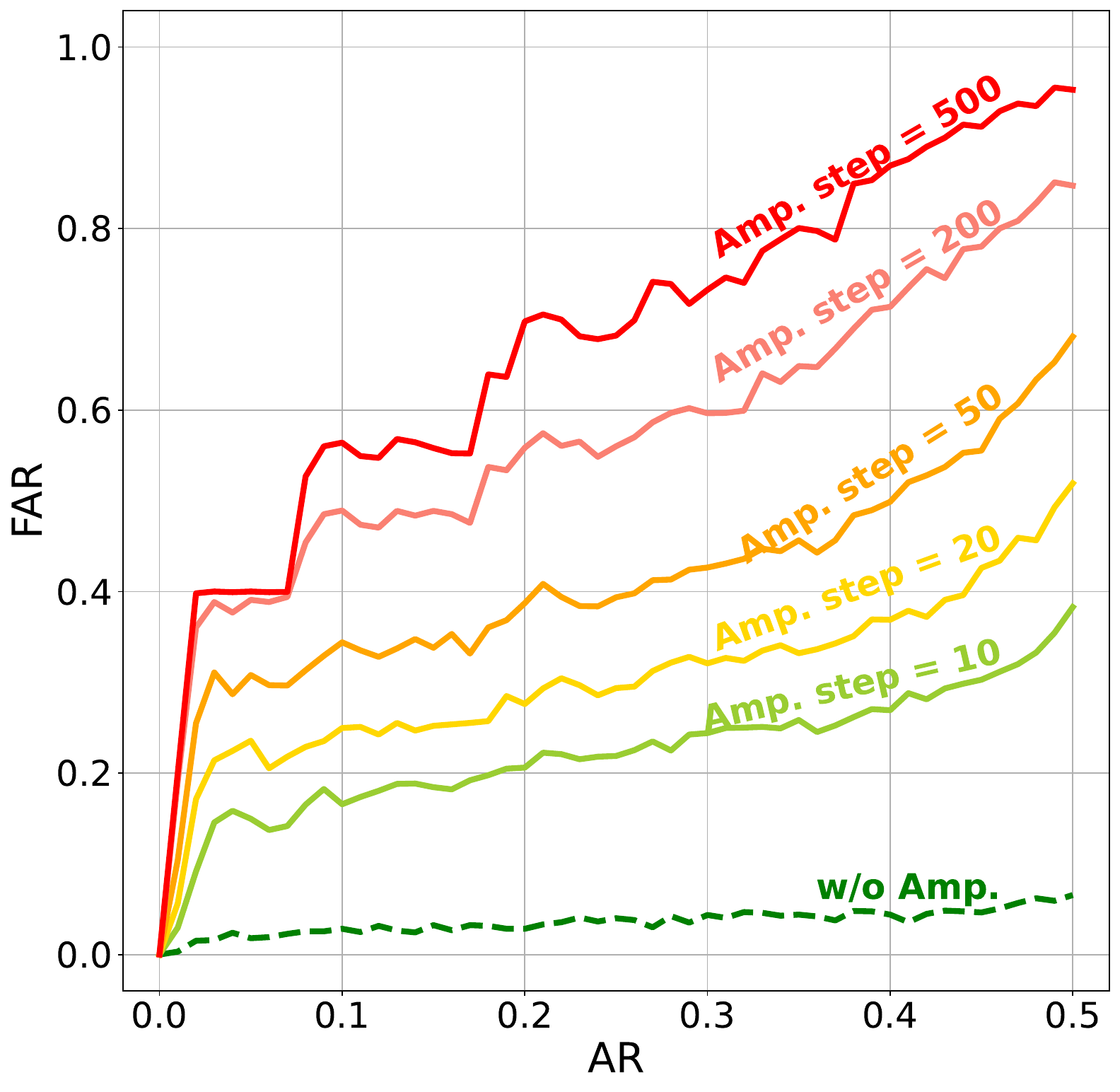}
        \caption{FAR-AR of Anti-NAF with different Amp. steps on LAION-mi.}
        \label{fig:laion_amp}
    \end{subfigure}
    \caption{The overall FAR-AR curves. The results show that \ourmethod is strongly effective in amplifying the possibility of infringed output, even with small \ourmethod steps.}
    \label{fig:res_plot}
\end{figure}

\subsection{Results}
\label{sec:results}
In \cref{fig:example}, we show example outputs under four scenarios.
It is evident that with \ourmethod, the CP-$k$ protected generative model does indeed output copyright-infringed content with high probability, especially when using the Anti-NAF prompt.
These examples suggest that probabilistic safeguards against copyright infringement are vulnerable to the \ourmethod attack.
Additional visualizations of outputs on various copyrighted images are demonstrated in \cref{fig:selected_examples}.
In \cref{tab:main_res}, we report CIRs and FARs at various ARs on two datasets.
Notably, there is a significant growth in both metrics when \ourmethod attack is employed, highlighting its effectiveness in amplifying the probability of infringing generations.
The superior performance of Anti-NAF underscores its efficacy in rendering infringing generations with a substantial probability.
The overall FAR-AR curves are illustrated in \cref{fig:res_plot}.
We can observe that bandit variants of \ourmethod lead to a smaller variance across different target copyrighted images, especially at lower acceptance rates, indicating that our bandit strategies achieve a more steady attack.
In \cref{fig:pokemon_amp,fig:laion_amp}, we plot the overall FAR-AR curves over different \ourmethod steps.
There is a clear trend of rapidly improved performance with increased \ourmethod steps, due to the cumulative probability of generating infringing samples.
This finding underlines the potential risk in practical applications of probabilistic copyright protections, given the high frequency of daily interactions with text-to-image generative models. 
We provide additional results and human evaluations in \cref{sec:detail_results}.

\subsection{Ablation Study}

In \cref{tab:ablation_pokemon}, we study how different components of our optimization objective in \cref{eq:loss}, affect the performance.
We discover that an exclusive focus on minimizing $\mathcal{L}_p$ (``$\mathcal{L}_p$ only'') leads to a notable drop in performance.
This indicates that prompts designed merely to reconstruct the target image are insufficient for a successful attack, as the infringing output of such a prompt can be easily identified and blocked by the copyright protection system.
Furthermore, a direct combination of $\mathcal{L}_q$ with $\mathcal{L}_p$ (``w/o $\varphi$'') results in additional performance degradation because of the conflicting objectives of $\mathcal{L}_p$ and $\mathcal{L}_q$ in minimizing $\rho(y_C|x)$.
On the other hand, we can observe a slight performance improvement when implementing a loss clip $\varphi$ on $\mathcal{L}_p$ (``w/o $\mathcal{L}_q$'') to constrain the learning of $\mathcal{L}_p$.
Overall, these results validate the effectiveness of our well-balanced optimization objective.
Additional ablation studies are provided in \cref{sec:add_ablation}.

\begin{table}[t]\footnotesize
    \centering
    \begin{tabular}{c|c|cc}
         \toprule
        Methods & CIR & FAR@5\%AR$\uparrow$ &  FAR@15\%AR$\uparrow$  \\
        \midrule
         Anti-NAF &	12.88\% & 8.52\% &	10.00\% \\
         \hline
         $\mathcal{L}_p$ only &	10.60\% & 1.84\% & 2.04\% \\
         w/o $\varphi$ &12.12\%	& 1.36\%	& 1.48\%	\\
         w/o $\mathcal{L}_q$ &	13.72\% & 2.00\%	& 2.76\%	\\
         
         \bottomrule
    \end{tabular}
    \caption{Ablation study for Anti-NAF algorithm on POKEMON.}
    \label{tab:ablation_pokemon}
\end{table}




%% file: sec/6_conclusion.tex
\section{Conclusion}
\label{sec:conclusion}
In this paper, we shed light on the vulnerability of probabilistic copyright protection methods for text-to-image generative models, especially in real-world scenarios involving persistent and targeted interactions.
Our proposed Virtually Assured Amplification Attack (VA3) framework presents the feasibility of inducing the protected model to generate copyright-infringed content with an amplified probability.
Despite our focus on a narrow scenario of online prompt selection within this framework, the experimental results highlight its effectiveness in challenging even the most advanced existing probabilistic copyright protection methods.
However, a broader scope of potential strategies remains unexplored within the VA3 framework, such as online prompt optimization, which may provide more powerful attacks against copyright protection.
Furthermore, our Anti-NAF algorithm relies on access to generative models for adversarial prompt optimization, an assumption that may not be satisfied in completely black-box scenarios.
We leave these more complicated and general attack methods for future investigation.
In conclusion, our findings emphasize the significant risk of copyright infringement when applying probabilistic copyright protection methods in practice.
Therefore, we hope that this work can inspire the development of more robust copyright protection approaches.

%% file: sec/7_acknowledgement.tex
\section*{Acknowledgement}
The authors are grateful to Nikhil Vyas for providing the source code of CP-$k$ algorithm.
This research is partially supported by the National Research Foundation Singapore under the AI Singapore Programme (AISG Award No: AISG2-TC-2023-010-SGIL) and the Singapore Ministry of Education Academic Research Fund Tier 1 (Award No: T1 251RES2207).

%% file: sec/X_suppl.tex
\clearpage
\maketitlesupplementary

\section{Proofs}

\begin{theorem_X}
    Following the notations in \cref{alg0}, for any $\varepsilon \in (0,1)$, the attack is successful with probability at least $1 - \varepsilon$ if $T > \log_{1-\sigma} \varepsilon$, where $\sigma > 0$ is a strictly positive lower-bound on the success probability shared by every single attack.
\end{theorem_X}

\begin{proof}
    Let $\mathcal{E}_t$ denote the event that the $t$-th attack is successful, and let $\mathcal{E}$ denote the event that at least one attack is successful.
    We want to proof that when $T > \log_{1-\sigma} \varepsilon$, 
    $$P(\mathcal{E}) > 1 - \varepsilon.$$
    The left-hand side of the inequality can be expanded as 
    \begin{equation}
        \begin{aligned}
            P(\mathcal{E}) =& P(\cup_{t=1}^T \mathcal{E}_t) = 1 - P(\cap_{t=1}^T \neg \mathcal{E}_t) \\
            =& 1 - \Pi_{t=1}^T P(\neg \mathcal{E}_t | \cap_{s=1}^{t-1} \neg\mathcal{E}_s) \\
            =& 1 - \Pi_{t=1}^T (1 - P(\mathcal{E}_t | \cap_{s=1}^{t-1} \neg\mathcal{E}_s))
        \end{aligned}
        \nonumber
    \end{equation}
    For every single attack we have a strictly positive lower-bound on the success probability, regardless of previous attacks.
    Specifically, we have $P(\mathcal{E}_t | \cap_{s=1}^{t-1} \neg\mathcal{E}_s) > \sigma$ for $t=1,\cdots,T$.
    Further considering $T > \log_{1-\sigma} \varepsilon$, we have
    \begin{equation}
        \begin{aligned}
            P(\mathcal{E}) =& 1 - \Pi_{t=1}^T (1 - P(\mathcal{E}_t | \cap_{s=1}^{t-1} \neg\mathcal{E}_s))\\
            \ge& 1 - (1 - \sigma)^T > 1 - \varepsilon
        \end{aligned}
        \nonumber
    \end{equation}

\end{proof}

\noindent We make the following side comment to avoid potential ambiguities in the statement of the theorem.
The statement \textit{A strictly positive lower-bound on the success probability shared by every single attack} DOES NOT refer to \textit{a strictly positive lower-bound on the marginal success probability} $P(\mathcal{E}_t), t=1,\cdots,T$.
It is obvious that $P(\mathcal{E}_t) > \sigma, t=1,\cdots,T$ do not lead to the conclusion, by considering the counter case where $P(\mathcal{E}_t | \cap_{s=1}^{t-1} \neg\mathcal{E}_s)=0, t=1,\cdots,T$.
The statement in the theorem is stronger, in the sense that the strictly positive lower-bound applies to the success probability of the attack at every step, regardless of previous attacks.
Or in other words, considering all possible previous attacks and results, the strictly positive lower-bound applies to the worst-case attack at the current step.

\begin{theorem_X}
    Assume there is a distance measure $\mathcal{D}$ defined in $\mathcal{Y}$ such that (i) $p$ is $(\epsilon_p, \alpha)$-local-continuous around $y_C$, (ii) every $q \in \mathcal{S}$ is local-continuous around $y_C$, and (iii) there exists $\epsilon_c > 0$ such that $\mathcal{B}_\mathcal{D}(y_C, \epsilon_c) \subseteq \mathcal{Y}_C$. The objective defined in \cref{eq:far} has the following lower-bound for any $\eta, \delta > 0$,
    \begin{equation}\nonumber
    \max_{x\in\mathcal{X}} P_{y\sim \tilde{p}(\cdot|x)}(y\in\mathcal{Y}_C)\\
    \ge \max_{x\in\tilde{\mathcal{X}}_{\eta, \delta}} \ \eta C_1 - \alpha C_2
    \end{equation}
    where $\tilde{\mathcal{X}}_{\eta, \delta} = \{x\in\mathcal{X}: p(y_C|x) \ge \eta, \rho(y_C|x) < k_x - \delta\}$ and $C_1, C_2$ are constants independent on $x$ given as
    \begin{equation}
        \begin{aligned}
            C_1 = \int_{y\in \mathcal{B}_{\mathcal{D}}(y_C, \epsilon)} dy, \ \ C_2 = \int_{y\in \mathcal{B}_{\mathcal{D}}(y_C, \epsilon)} \mathcal{D}(y_C, y)dy,
        \end{aligned}
        \nonumber
    \end{equation}
    where $\epsilon = \min(\epsilon_p, \epsilon_c, \epsilon_\rho)$ with $\epsilon_\rho:={\rm inf}_{x\in\tilde{\mathcal{X}}_{\eta, \delta}}{\rm sup}\{\epsilon: \rho(y|x) < k_x, \forall y \in \mathcal{B}_{\mathcal{D}}(y_C, \epsilon)\}.$
\end{theorem_X}

\begin{proof} First, let us prove $\epsilon > 0$.
$\epsilon_p > 0$ and $\epsilon_c >0$ are assured by the assumptions, so we only need to prove $\epsilon_\rho > 0$.
As $p$ and every $q \in \mathcal{S}$ are assumed to be local-continuous around $y_C$, $\rho$ is local-continuous around $y_C$, say $(\tilde{\epsilon}, \beta)$-local-continuous.
For any $x \in \tilde{\mathcal{X}}_{\eta, \delta}$ and $y \in \mathcal{B}_{\mathcal{D}}(y_C, \tilde{\epsilon})$, we have $|\rho(y_C|x) - \rho(y|x)| < \beta \mathcal{D}(y_C, y)$.
Further,
\begin{equation}\nonumber
    \begin{aligned}
        \rho(y|x) \le& \rho(y_C|x) + |\rho(y_C|x) - \rho(y|x)|\\
        <& k_x - \delta + \beta \mathcal{D}(y_C, y)\\
    \end{aligned}
\end{equation}
For $y\in\mathcal{B}(y_C, \min(\tilde{\epsilon}, \delta/\beta))$, $\rho(y|x) < k_x$.
Thus, $\epsilon_\rho \ge \min(\tilde{\epsilon}, \delta/\beta) > 0$.

Next, let us move back to the main objective. 
By applying Bayes' theorem, we have
\begin{equation}
\begin{aligned}
    &\max_{x\in\mathcal{X}} P_{y\sim \tilde{p}(\cdot|x)}(y\in\mathcal{Y}_C)\\
    =& \max_{x\in\mathcal{X}}P_{y\sim p(\cdot|x)}(y\in\mathcal{Y}_C|\rho(y|x) < k_x)\\
    =& \max_{x\in\mathcal{X}} \frac{P_{y\sim p(\cdot|x)}(\rho(y|x) < k_x, y\in\mathcal{Y}_C)}{P_{y\sim p(\cdot|x)}(\rho(y|x) < k_x)}\\
    \ge& \max_{x\in\mathcal{X}} P_{y\sim p(\cdot|x)}(\rho(y|x) < k_x, y\in\mathcal{Y}_C)\\
    =& \max_{x\in\mathcal{X}}\int_{y\in \mathcal{Y}} \mathbb{I}(y\in\mathcal{Y}_C)\mathbb{I}(\rho(y|x) < k_x) p(y|x)dy.
\nonumber
\end{aligned}
\end{equation}
The inequality comes from $P_{y\sim p(\cdot|x)}(\rho(y|x) < k_x) \le 1$.
We will next only consider prompts in $\tilde{\mathcal{X}}_{\eta, \delta}$.
Recall that for any $x \in \tilde{\mathcal{X}}_{\eta, \delta}$ and $y \in \mathcal{B}_\mathcal{D}(y_C, \epsilon)$, we have $y \in \mathcal{Y}_C$ and $\rho(y|x) < k_x$.
Thus, we can remove the two indicators by narrowing the scope of integral to $\mathcal{B}_\mathcal{D}(y_C, \epsilon)$.

\begin{equation}
\begin{aligned}
    &\max_{x\in\mathcal{X}} P_{y\sim \tilde{p}(\cdot|x)}(y\in\mathcal{Y}_C)\\
    \ge& \max_{x\in\mathcal{X}}\int_{y\in \mathcal{Y}} \mathbb{I}(y\in\mathcal{Y}_C)\mathbb{I}(\rho(y|x) < k_x) p(y|x)dy\\
    \ge& \max_{x\in\tilde{\mathcal{X}}_\eta}\int_{y\in \mathcal{B}_\mathcal{D}(y_C, \epsilon)} \mathbb{I}(y\in\mathcal{Y}_C)\mathbb{I}(\rho(y|x) < k_x) p(y|x)dy\\
    =& \max_{x\in\tilde{\mathcal{X}}_\eta}\int_{y\in \mathcal{B}_\mathcal{D}(y_C, \epsilon)} p(y|x)dy\\
\nonumber
\end{aligned}
\end{equation}
Finally, by utilizing the local-continuity of $p$, we get the desired lower-bound.
\begin{equation}
\begin{aligned}
    &\max_{x\in\mathcal{X}} P_{y\sim \tilde{p}(\cdot|x)}(y\in\mathcal{Y}_C)\\
    \ge& \max_{x\in\tilde{\mathcal{X}}_\eta}\int_{y\in \mathcal{B}_{\mathcal{D}}(y_C, \epsilon)} p(y|x)dy\\
    \ge& \max_{x\in\tilde{\mathcal{X}}_\eta}\int_{y\in \mathcal{B}_{\mathcal{D}}(y_C, \epsilon)} [p(y_C|x) - \alpha \mathcal{D}(y_C, y)]dy\\
    \ge& \max_{x\in\tilde{\mathcal{X}}_\eta}\int_{y\in \mathcal{B}_{\mathcal{D}}(y_C, \epsilon)} [\eta - \alpha \mathcal{D}(y_C, y)]dy\\
    =& \ \max_{x\in\tilde{\mathcal{X}}_\eta} \eta C_1 - \alpha C_2
\nonumber
\end{aligned}
\end{equation}

\end{proof}

\section{On Future work}
In this paper, we consider the setting where an attacker can interact with a target model in the online manner. Future work includes the setting of transferring an attack from a set of source models to a target model via a generalization property of attacks \citep{zou2023universal} by controlling the mutual information to avoid over-fitting to the source models \citep{icml2023kzxinfodl}.

\section{Details on Fine-tuning}
We fine-tune the pre-trained StableDiffusion-v1-4 model provided by Huggingface on two datasets.
Given the different sizes of the two datasets, the fine-tuning steps are set to 5000 and 25,000 for POKEMON and LAION-mi respectively.
For fine-tuning both datasets, the batch size is set to 1, the gradient accumulations step is set to 4, and the learning rate is 1e-5.

\section{Infringement Judgment}
\label{sec:infringe}
\begin{figure*}[htbp]
    \begin{subfigure}[b]{\textwidth}
        \includegraphics[width=\textwidth]{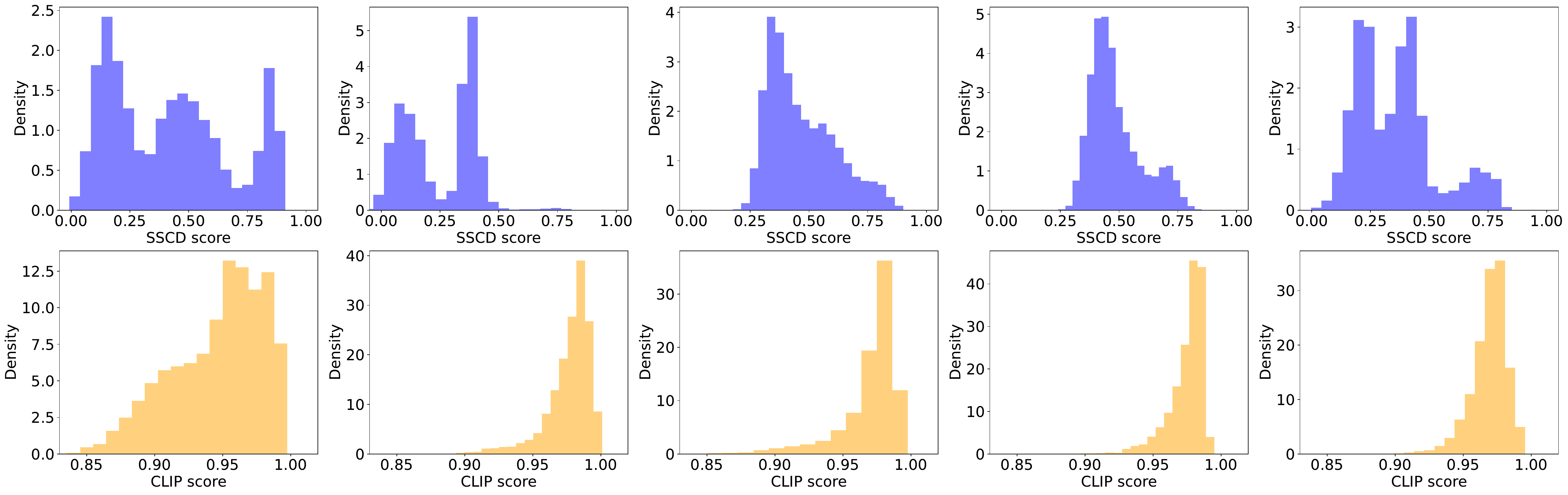}
        \caption{Distributions of SSCD and CLIP score on 5 copyrighted images in the POKEMON dataset. Each column corresponds to one target image.}
        \label{fig:pokemon_sim}
    \end{subfigure}
    \begin{subfigure}[b]{\textwidth}
        \includegraphics[width=\textwidth]{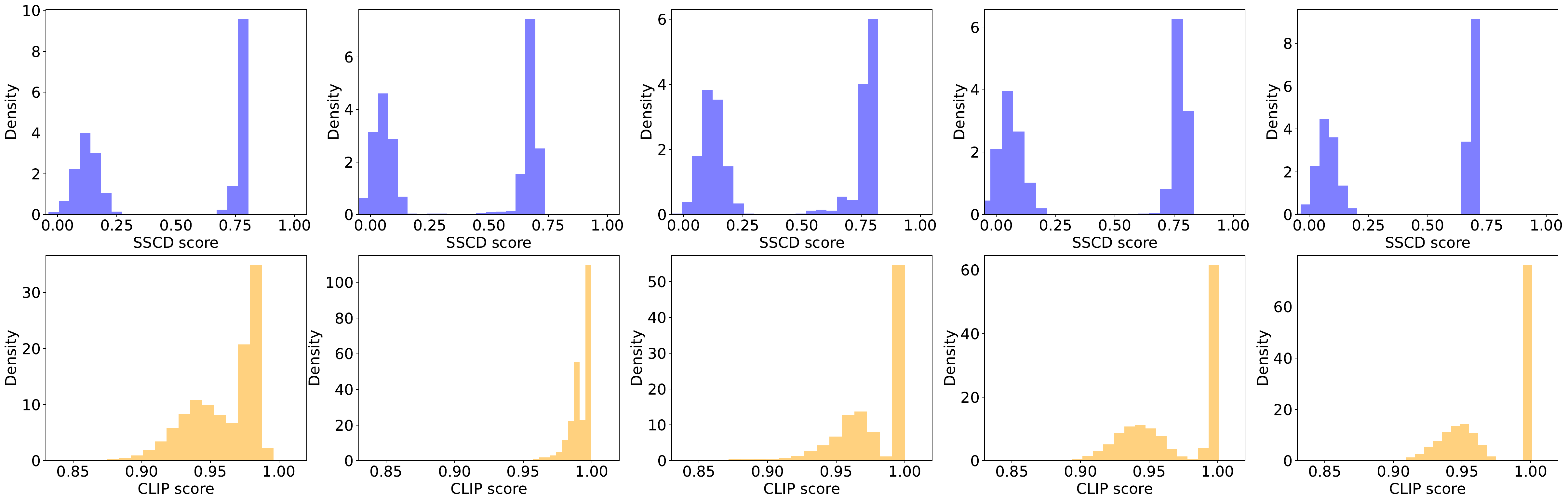}
        \caption{Distributions of SSCD and CLIP score on 5 copyrighted images in the LAION-mi dataset. Each column corresponds to one target image.}
        \label{fig:laion_sim}
    \end{subfigure}
    \caption{Distributions of SSCD and CLIP similarity score on all target copyrighted images in two datasets using the original caption as prompts. The distributions of the SSCD score are more clearly bimodal to distinguish between non-infringing and infringing samples.}
    \label{fig:sim_comp}
\end{figure*}

\begin{figure*}[htbp]
    \centering
    \includegraphics[width=\textwidth]{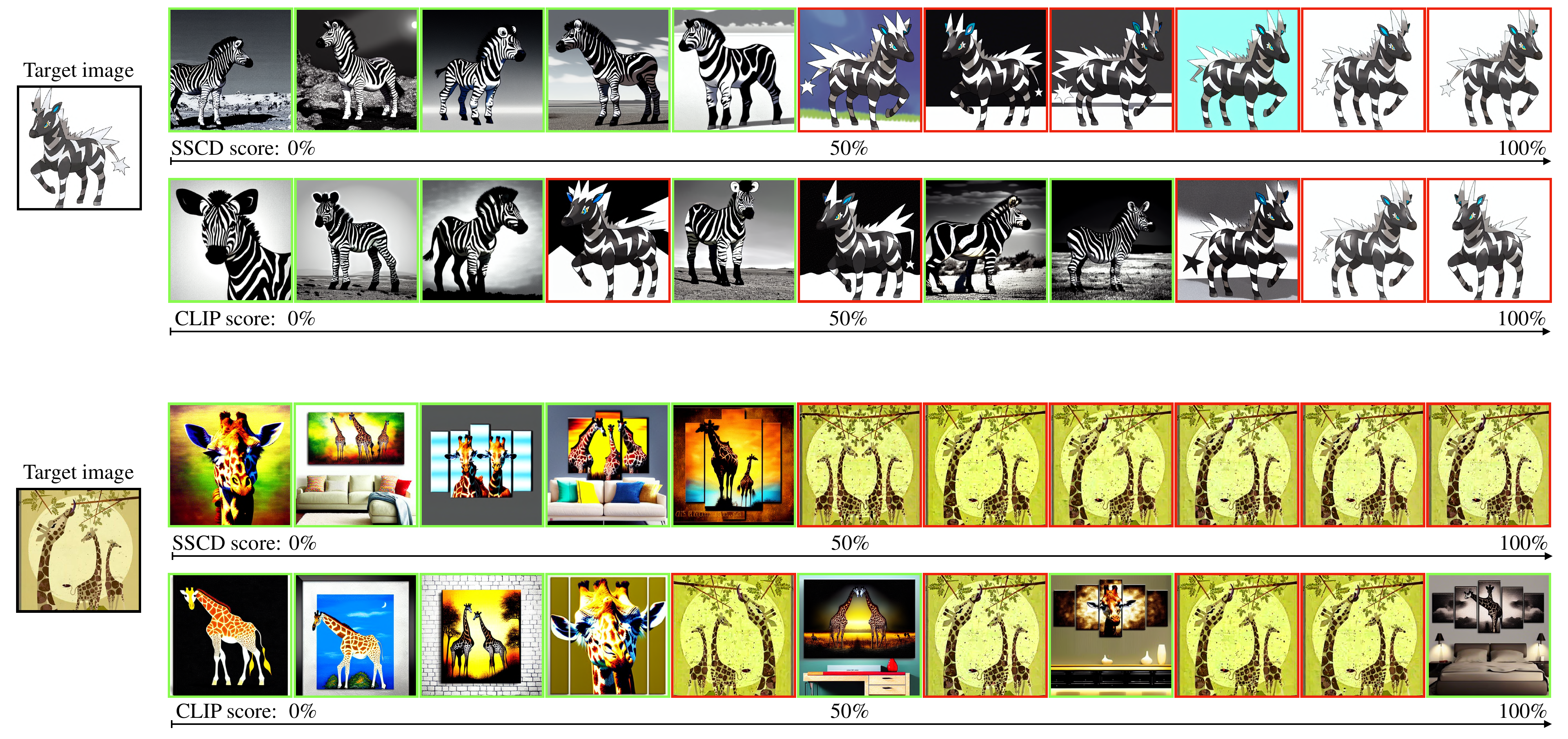}
    \caption{Example images generated from the original caption of target images (non-infringing and infringing images are marked with green and red boundaries, respectively). From left to right, images are sorted by similarity score in ascending order. An ideal similarity score threshold should distinguish between non-infringing (lower score) and infringing samples (higher score). From the example images, the SSCD score performs much better than the CLIP score.}
    \label{fig:sim_example}
\end{figure*}

To determine whether samples generated by model $p$ infringe the copyright of the target image, we need to assign ground-truth labels to these samples.
Unfortunately, to our knowledge, there is currently no widely recognized computable standard for determining whether an image infringes copyright. 
In fact, the criteria for copyright infringement determination may evolve with changing societal perceptions.
Alternatively, we rely on the similarity between samples and the target image as the basis for determining infringement. 
In order to distinguish between non-infringing and infringing samples, an ideal similarity score should assign lower scores to non-infringing samples and higher scores to infringing samples.
Recognizing the limitations of a singular similarity measure, we compare the performance of SSCD \cite{pizzi2022self} and CLIP score for determining copyright infringement.
In \cref{fig:sim_comp}, we plot the histograms of SSCD scores and CLIP scores for images generated by original captions of all target copyrighted images in two datasets.
We can observe that the distributions of SSCD scores demonstrate a more clearly bimodal pattern compared with CLIP scores.
This means that non-infringing and infringing samples can be better distinguished by the two modes of distribution of SSCD scores.
In \cref{fig:sim_example}, we show example images with different values of similarity scores in ascending order.
We can find that non-infringing samples may have higher CLIP scores than infringing samples with target images.
However, there is a clear threshold (\eg, 50\%) for SSCD score to distinguish non-infringing and infringing samples.
Thus, in this paper, we use the SSCD score for infringement judgment.

In \cref{sec:results}, we report results with SSCD-50\% as the infringement threshold.
Further, considering the evolving nature of copyright infringement standards, we utilize other varying thresholds.
According to the observation in \cref{fig:sim_comp}, we consider modeling the SSCD score distribution as a mixture of two Gaussian distributions and use the mean value of two means of the Gaussian distributions as the similarity threshold, denoted as SSCD-gmm.
For POKEMON dataset, we further consider SSCD-45\% and SSCD-55\%.

\begin{figure}[t]
    \centering
    \includegraphics[width=\linewidth]{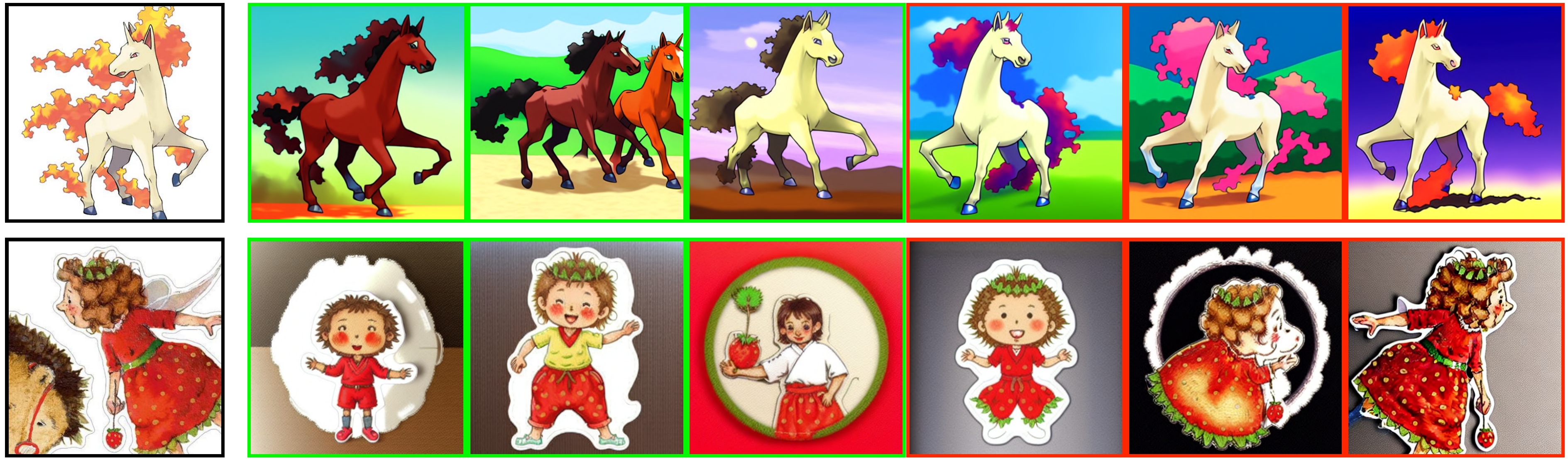}
    \caption{Qualitative examples near the similarity decision thresholds (target, non-infringing, and infringing images are marked with black, green, and red boundaries, respectively).}
    \label{fig:boundary_examples}
\end{figure}

In \cref{fig:boundary_examples}, we also provide qualitative examples that are close to the decision thresholds using Anti-NAF prompts for generation.
The qualitative examples verify that similarity decision thresholds can be utilized to clarify between style-similar and copyright-infringed generations effectively.

\section{Results}
\label{sec:detail_results}
In this section, we provide a detailed analysis of results in \cref{sec:results} and additional results on human evaluation, other similarity thresholds, and transfer attack.

\subsection{Detailed Analysis on Results}
\begin{figure*}[htbp]
    \centering
    \includegraphics[width=\textwidth]{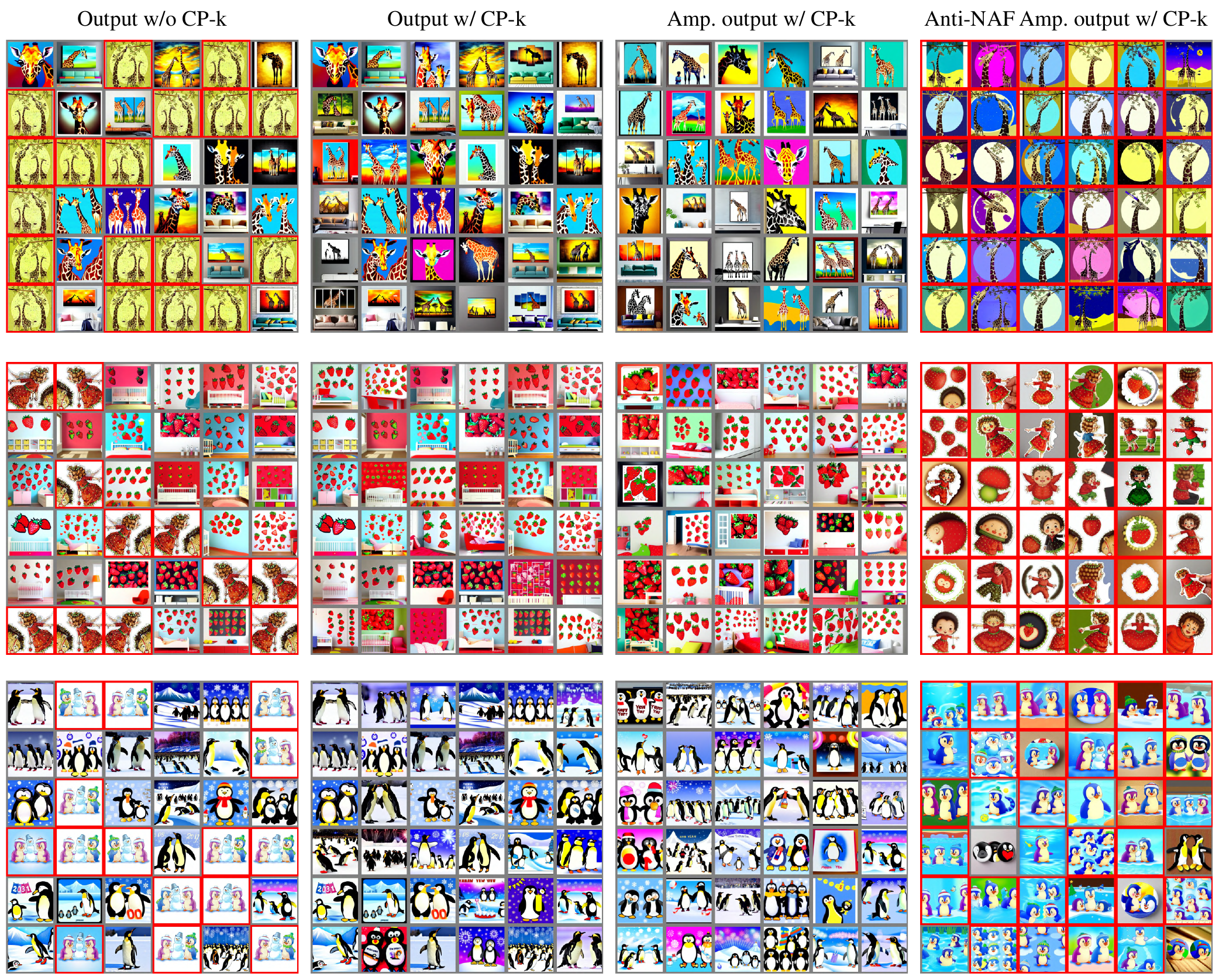}
    \caption{Example outputs given the copyright images in the second row of \cref{fig:selected_examples} as targets (potential infringing images are marked with red boundaries).
    In the first column, using a benign prompt, we observe a high incidence of infringing content from models without copyright protection (``w/o CP-$k$'').
    In contrast, all samples in the second column are safe after applying the copyright protection mechanism (``w/ CP-$k$'').
    In the third column, we find that amplification (Amp.) attack with a benign prompt can be unsuccessful.
    However, by amplification attack with an adversarial prompt obtained from our proposed Anti-NAF algorithm, most outputs in the last column are copyright-infringed.}
    \label{fig:app_example}
\end{figure*}

In \cref{fig:app_example}, we show example outputs of three target copyrighted images under four attack and defense scenarios. 
Similar to \cref{fig:example}, using a benign prompt (such as the original caption), in the first column, we can observe that outputs without copyright protection infringe the copyright of target images with high probability; in the second column, after copyright protection, all samples are non-infringing content as CP-$k$ rejects all infringing samples.
In the third column, we find that an amplification attack with a benign prompt can be unsuccessful, because such a prompt may not provide a strictly positive probability of producing infringing generations from models protected by CP-$k$.
However, in the last column, with an adversarial prompt obtained from our proposed Anti-NAF algorithm, we can see that most of the outputs are copyright-infringed, which means that the probability of infringing samples is largely amplified. 

\begin{figure*}[htbp]
    \centering
    \begin{subfigure}[b]{0.3\textwidth}
        \includegraphics[width=\textwidth]{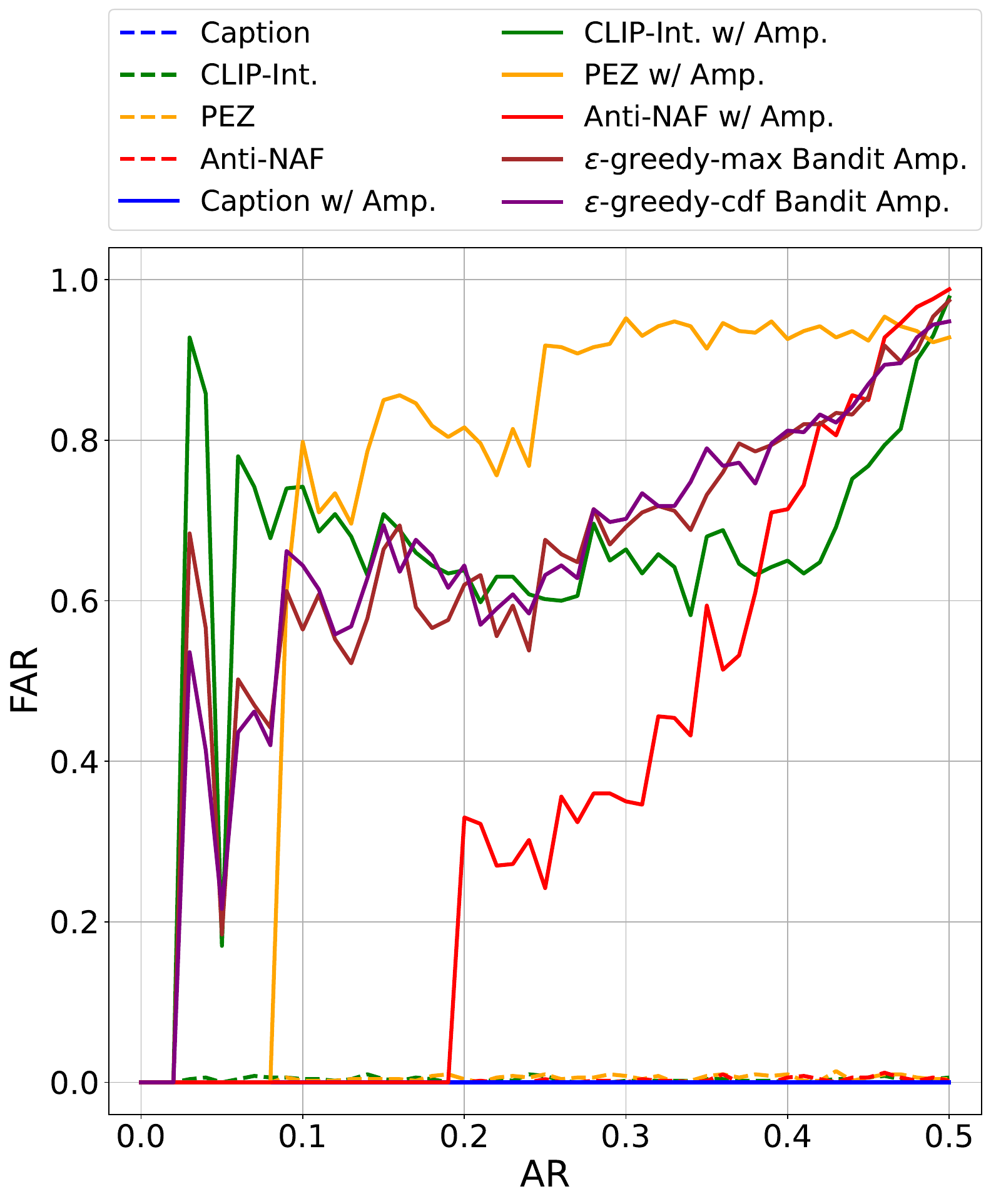}
        \caption{FAR-AR curves on No.1 target.}
        \label{fig:laion_far_ar_1}
    \end{subfigure}
    \hfill
    \begin{subfigure}[b]{0.3\textwidth}
        \includegraphics[width=\textwidth]{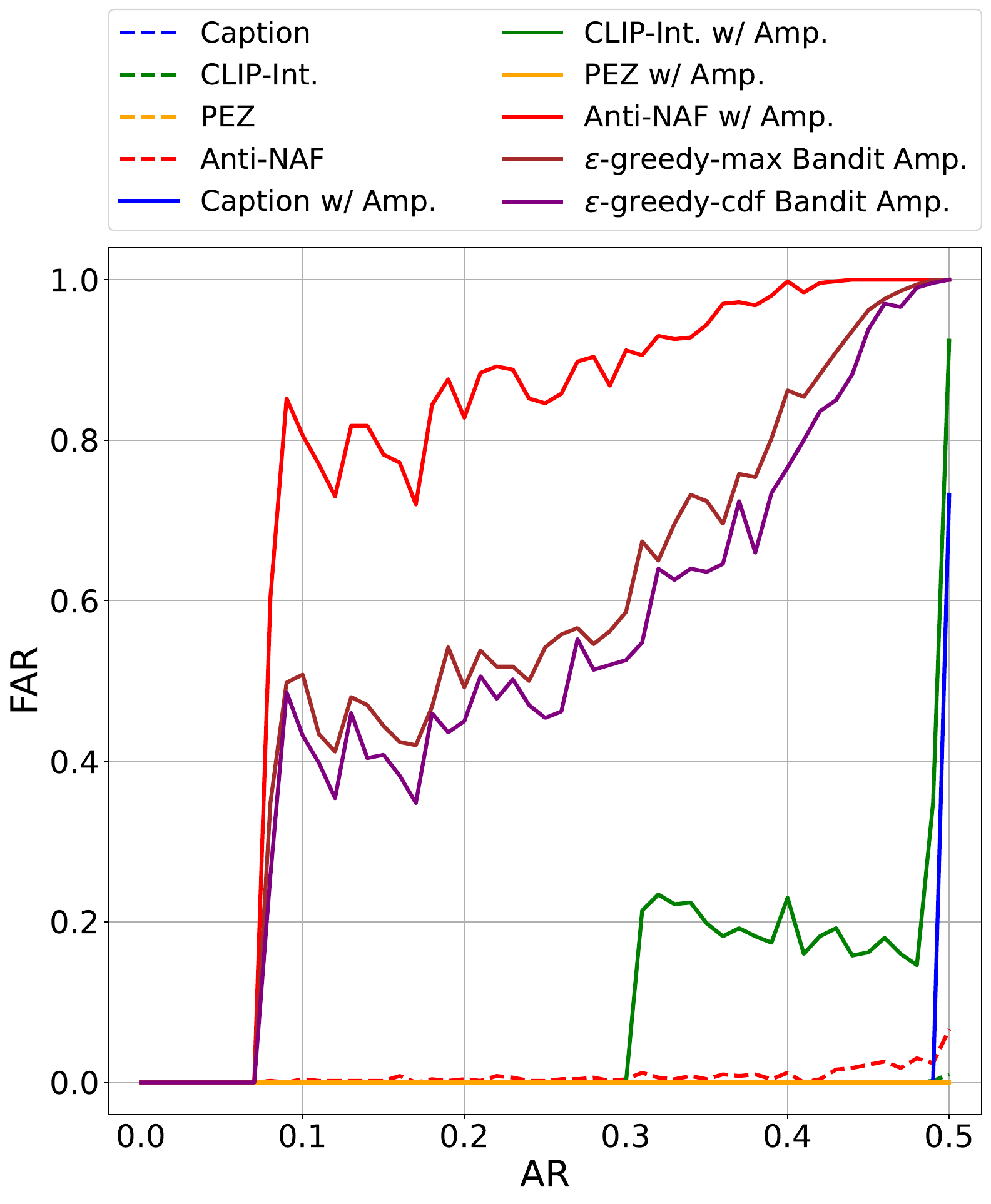}
        \caption{FAR-AR curves on No.2 target.}
        \label{fig:laion_far_ar_2}
    \end{subfigure}
    \hfill
    \begin{subfigure}[b]{0.3\textwidth}
        \includegraphics[width=\textwidth]{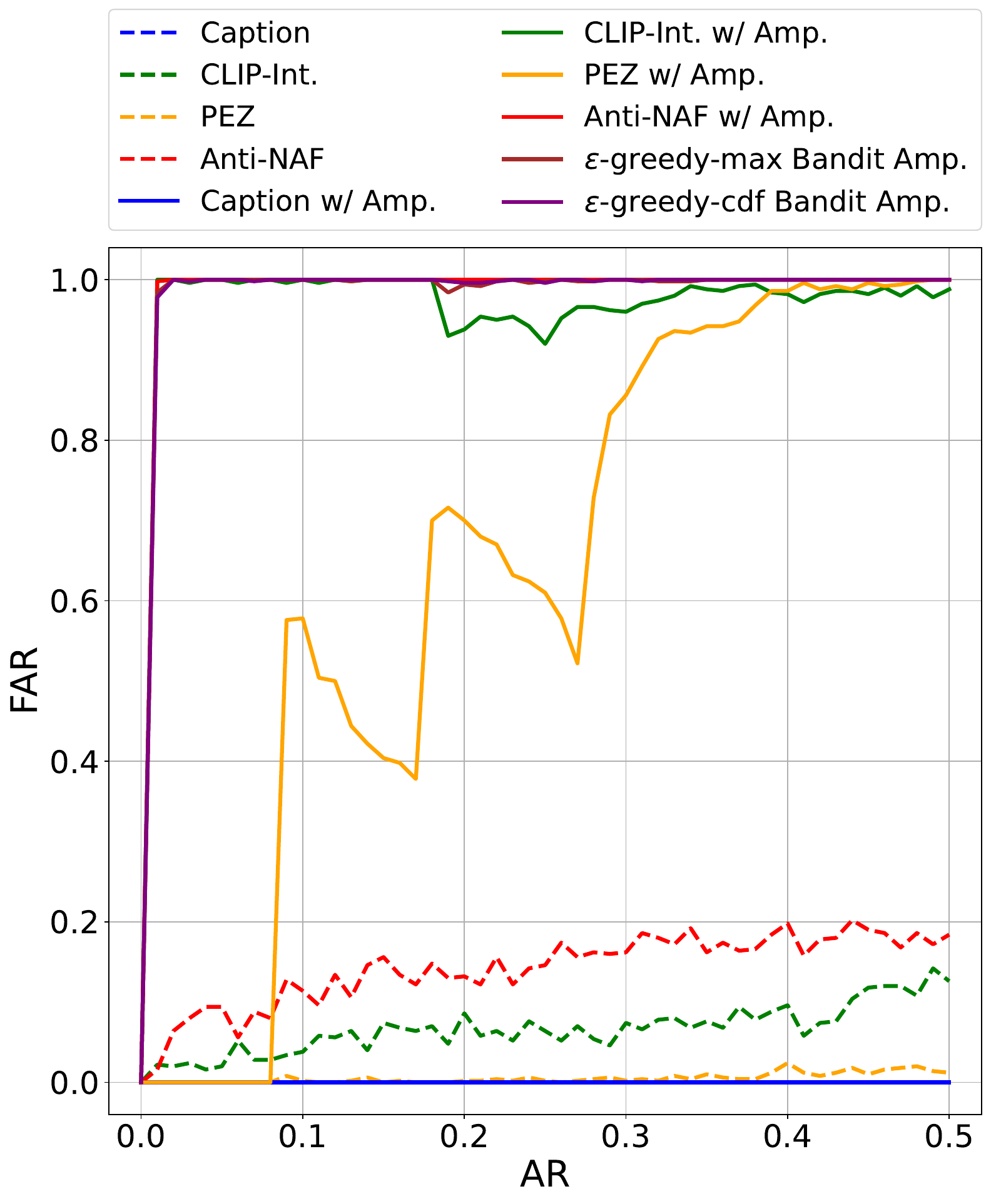}
        \caption{FAR-AR curves on No.3 target.}
        \label{laion_far_ar_3}
    \end{subfigure}
    \hfill
    \begin{subfigure}[b]{0.3\textwidth}
        \includegraphics[width=\textwidth]{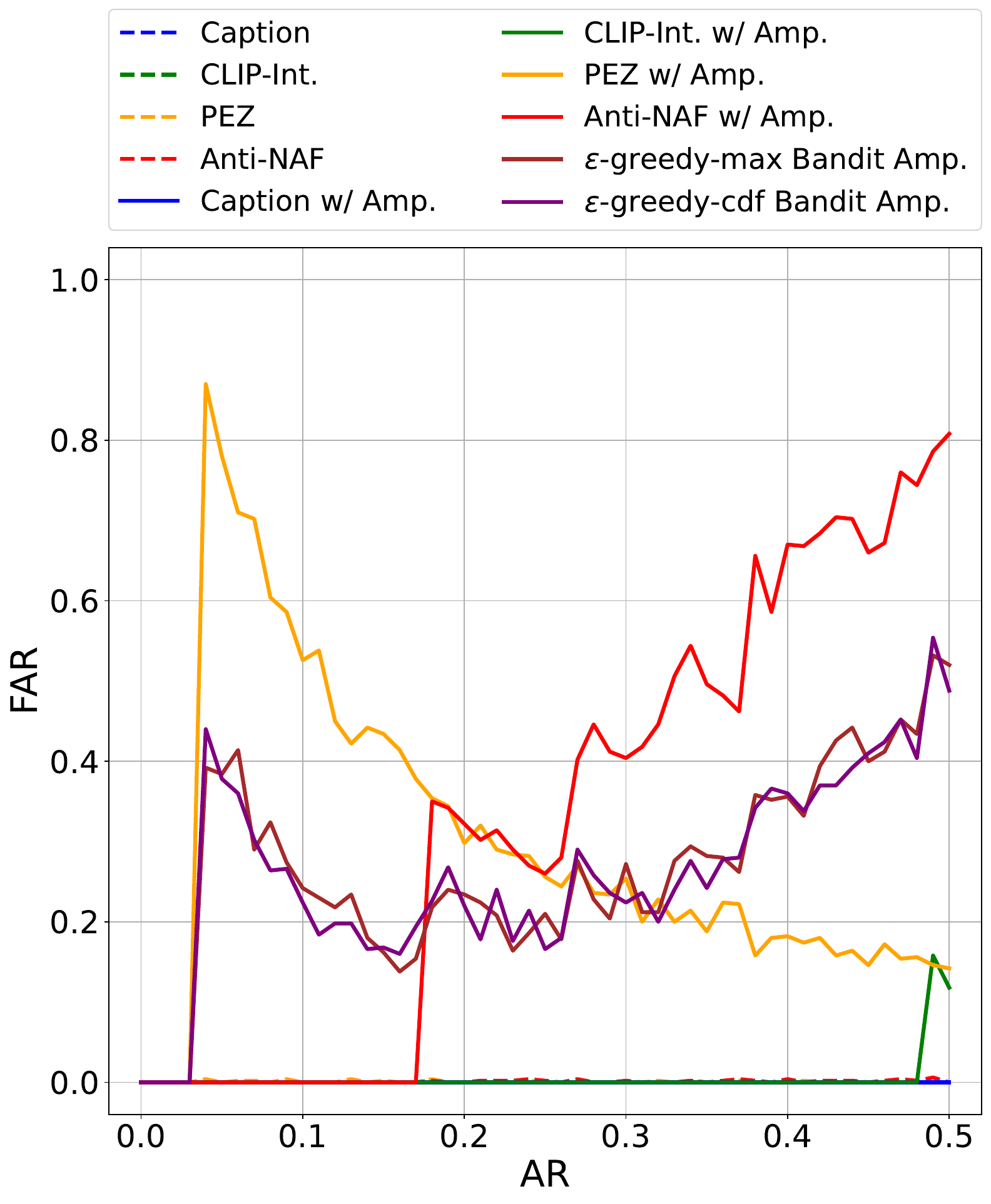}
        \caption{FAR-AR curves on No.4 target.}
        \label{fig:laion_far_ar_4}
    \end{subfigure}
    \hfill
    \begin{subfigure}[b]{0.3\textwidth}
        \includegraphics[width=\textwidth]{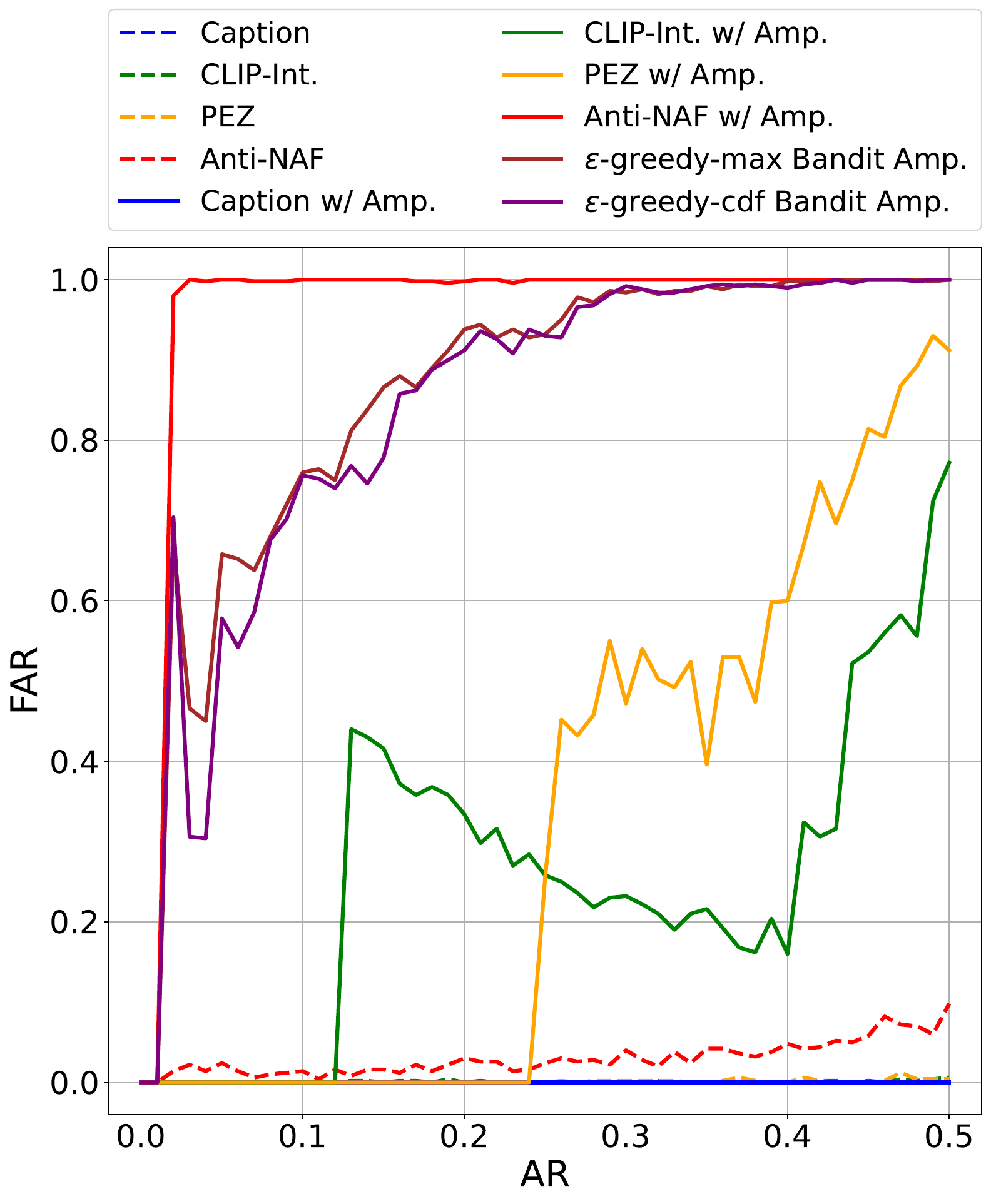}
        \caption{FAR-AR curves on No.5 target.}
        \label{fig:laion_far_ar_5}
    \end{subfigure}
    \hfill
    \begin{subfigure}[b]{0.3\textwidth}
        \includegraphics[width=\textwidth]{fig/FAR_AR_laion_avg5_boost500_v2.pdf}
        \caption{Overall averaged FAR-AR curves.}
        \label{fig:laion_far_ar_avg}
    \end{subfigure}
    \caption{FAR-AR curves on each copyrighted image in LAION-mi.
    For No.1 and 4 target copyrighted images, Anti-NAF performs worse when acceptance rate is lower than 20\%, while bandit amplification methods show steady performance in these worst cases.}
    \label{fig:laion_far_ar}
\end{figure*}

In \cref{fig:laion_far_ar}, we give detailed FAR-AR curves on each target copyrighted image in LAION-mi dataset. 
We can find that our proposed bandit amplification method performs more steadily in the worst cases.
For example, in \cref{fig:laion_far_ar_1,fig:laion_far_ar_4}, when acceptance rate is lower than 20\%, the FAR of Anti-NAF with amplification is nearly 0\%; while $\varepsilon$-greedy-max/-cdf bandit amplification can adapt to follow the best choice of prompts (\eg, PEZ or CLIP-Interrogator) and keep a competitive FAR score.

\subsection{Human Evaluation}
In \cref{tab:human_eval}, we conduct a human evaluation on two target copyrighted images from two datasets. 
We randomly select 100 accepted samples obtained from each of the two threat models (the original caption and $\varepsilon$-greedy-cdf).
For each target image, a total of 200 samples are randomly shuffled and displayed to 5 graduate students. They are told to label each sample as non-infringing or infringing the copyright of the given target image.
Finally, we report their average copyright infringement rates.

\subsection{Results on Other Similarity Thresholds}
\label{sec:add_results}
The results on the additional thresholds described in \cref{sec:infringe} are reported in \cref{tab:gmm_res,tab:pokemon_res}. 
We can find that under more strict similarity thresholds, our proposed Anti-NAF can also provide a non-trivial probability of producing infringing content even with a low acceptance rate. 
Besides, Anti-NAF outperforms other threat prompts under all different similarity thresholds, highlighting its effectiveness.

\subsection{Results on Transfer Attack}
In \cref{tab:pokemon_sd-1-5}, we investigate the generalizability of our proposed Anti-NAF algorithms on transfer attack settings.
Specifically, the adversarial prompt optimization is conducted based on a fine-tuned StableDiffusion-v1-4 model, while the obtained prompts are then utilized to attack the fine-tuned StableDiffusion-v1-5 model.
The results indicate that prompts generated on a white-box model using Anti-NAF can serve as candidate prompts for VA3 to attack other black-box models.
We hope this study can inspire future work to explore black-box attacks in practical scenarios.

\begin{table}[htbp]\footnotesize
    \centering
    \begin{tabular}{c|cc}
         \toprule
        Dataset & Caption (w/o Amp.)  &  $\varepsilon$-greedy-cdf Amp.  \\
        \midrule
        POKEMON & 0.6\% & 83.0\%\\
        LAION-mi & 0.4\% & 42.4\%\\
         \bottomrule
    \end{tabular}
    \caption{Human evaluation results of copyright-infringement rate on selected target images of two datasets. Acceptance rates of 10\% and 40\% are applied for POKEMON and LAION-mi respectively.}
    \label{tab:human_eval}
\end{table}

\begin{table}[htbp]\footnotesize
    \centering
    \begin{tabular}{c|cc}
       \toprule
         Methods & FAR@5\%AR$\uparrow$ &  FAR@15\%AR$\uparrow$ \\
        \midrule
        Caption & 2.13\% & 11.07\%\\
         Anti-NAF &	\textbf{9.07\%}	& \textbf{21.87\%}  \\
         \bottomrule
    \end{tabular}
    \caption{Results on selected target images of POKEMON. The prompts of Anti-NAF are obtained with StableDiffusion-v1-4, while attacks are conducted on StableDiffusion-v1-5.}
    \label{tab:pokemon_sd-1-5}
\end{table}

\section{Additional Results on Ablation Study}
\label{sec:add_ablation}
In \cref{tab:ablation_laion}, we report the results of the ablation study on LAION-mi dataset. 
We can observe that the results show similar trends as that of the POKEMON dataset in \cref{tab:ablation_pokemon}.
This further verifies that the optimization objective of our proposed Anti-NAF algorithm is effective and well-balanced between $\mathcal{L}_p$ and $\mathcal{L}_q$ with the help of loss clip bound $\varphi$.
In \cref{tab:pokemon_t25}, we also report ablation experiments on other choices of denoising steps $T$ of text-to-image diffusion models.
We can find that our proposed Anti-NAF keeps superior performance, suggesting that its effectiveness is immune to different $T$.

\begin{table}[htbp]\footnotesize
    \centering
    \begin{tabular}{c|c|cc}
       \toprule
         $T$ & Methods & FAR@5\%AR$\uparrow$ &  FAR@15\%AR$\uparrow$ \\
        \midrule
        \multirow{2}*{25} & Caption & 0.68\% & 3.52\%\\
         ~ & Anti-NAF &	\textbf{12.32\%}	& \textbf{14.44\%}  \\
         \hline
         \multirow{2}*{100} & Caption & 0.00\% & 3.40\%\\
         ~ & Anti-NAF &	\textbf{9.24\%}	& \textbf{9.52\%}  \\
         \bottomrule
    \end{tabular}
    \vspace{-0.3cm}
    \caption{Results with different denoising steps $T$ on POKEMON.}
    \label{tab:pokemon_t25}
\end{table}

\begin{table*}[htbp]\footnotesize
    \centering
    \begin{tabular}{l|c|cc|c|cc}
    \toprule
    \multirow{2}*{Methods} & \multicolumn{3}{c|}{SSCD-45\%} & \multicolumn{3}{c}{SSCD-55\%}\\
    \cline{2-7}
    ~ & CIR & FAR@5\%AR$\uparrow$ &FAR@15\%AR$\uparrow$ & CIR &  FAR@5\%AR$\uparrow$ & FAR@15\%AR$\uparrow$\\
    \midrule
    
    Caption (w/o \ourmethodshort)&	47.96\%&	0.84\%&	3.60\%&	42.64\%&	0.48\%&	2.64\% \\
    
    CLIP-Int. (w/o \ourmethodshort)& 31.28\%&	3.44\%&	5.24\%&	18.64\%&	0.64\%&	1.48\%\\
    
    PEZ (w/o \ourmethodshort)& 13.88\%&	3.28\%&	5.64\%&	5.60\%&	0.92\%&	1.52\%\\
    
    Anti-NAF (w/o \ourmethodshort)&	22.56\%&	\textbf{14.68\%}&	\textbf{19.80\%}&	8.08\%&	\textbf{5.08\%}&	\textbf{6.52\%} \\
    \cline{1-7}
    
    Caption (w/ \ourmethodshort)&	100.00\%&	14.64\%&	38.72\%&	100.00\%&	14.64\%&	38.68\% \\
    
    CLIP-Int. (w/ \ourmethodshort)&	99.84\%&	24.12\%&	48.00\%&	99.84\%&	17.64\%&	44.16\% \\
    
    PEZ (w/ \ourmethodshort)&	74.44\%&	30.64\%&	48.88\%&	63.32\%&	15.52\%&	34.28\%	\\
    
    Anti-NAF (w/ \ourmethodshort)&	99.92\%&	\textbf{86.28\%}&	\textbf{96.48\%}&	99.36\%&	\textbf{62.12\%}&	\textbf{66.44\%} \\
    
    \bottomrule
    \end{tabular}
    \caption{Quantitative results on POKEMON dataset using SSCD-45\% and SSCD-55\% as the threshold for infringement judgment. 
    (CLIP-Int. is the abbreviation for CLIP-Interrogator).}
    \label{tab:pokemon_res}
\end{table*}

\begin{table*}[htbp]\footnotesize
    \centering
    \begin{tabular}{l|c|cc|c|ccc}
    \toprule
    \multirow{2}*{Methods} & \multicolumn{3}{c|}{POKEMON} & \multicolumn{4}{c}{LAION-mi}\\
    \cline{2-8}
    ~ & CIR & FAR@5\%AR$\uparrow$ &FAR@15\%AR$\uparrow$ & CIR &  FAR@10\%AR$\uparrow$ & FAR@30\%AR$\uparrow$ & FAR@50\%AR$\uparrow$ \\
    \midrule
    
    Caption (w/o \ourmethodshort)&	40.40\%&	0.08\%&	1.52\%&	48.52\%&	0.00\%&	0.00\%&	0.08\% \\
    
    CLIP-Int. (w/o \ourmethodshort)& 23.96\%&	1.72\%&	2.76\%&	38.04\%&	0.00\%&	0.00\%&	0.20\%	\\
    
    PEZ (w/o \ourmethodshort)&8.28\%&	0.28\%&	0.80\%&	9.64\%&	0.00\%&	0.00\%&	0.00\%	\\
    
    Anti-NAF (w/o \ourmethodshort)&	16.20\%&	\textbf{11.48\%}&	\textbf{13.12\%}&	26.32\%&	\textbf{0.12\%}&	\textbf{0.20\%}&	\textbf{1.68\%} \\
    \cline{1-8}
    
    Caption (w/ \ourmethodshort)&	100.00\%&	3.56\%&	35.52\%&	100.00\%&	0.00\%&	0.00\%&	14.72\% \\
    
    CLIP-Int. (w/ \ourmethodshort)&	97.96\%&	14.68\%&	26.36\%&	100.00\%&	0.00\%&	0.00\%&	46.84\% \\
    
    PEZ (w/ \ourmethodshort)&	61.00\%&	10.80\%&	25.44\%&	81.56\%&	0.00\%&	0.00\%&	4.92\%	\\
    
    Anti-NAF (w/ \ourmethodshort)&	89.28\%&	\textbf{48.32\%}&	\textbf{67.04\%}&	99.68\%&	\textbf{26.24\%}&	\textbf{39.96\%}&	\textbf{59.44\%} \\
    
    
    
   
    \bottomrule
    \end{tabular}
    \caption{Quantitative results using SSCD-gmm as the threshold for infringement judgment. 
    (CLIP-Int. is the abbreviation for CLIP-Interrogator).}
    \label{tab:gmm_res}
\end{table*}

\begin{table*}[htbp]\footnotesize
    \centering
    \begin{tabular}{c|c|ccc}
         \toprule
        Methods & CIR & FAR@10\%AR$\uparrow$ &  FAR@30\%AR$\uparrow$ &  FAR@50\%AR$\uparrow$ \\
        \midrule
         Anti-NAF &	33.84\%	& 2.64\% & 	4.16\% & 7.00\% \\
         \hline
         $\mathcal{L}_p$ only &	30.44\%	& 0.28\% &	1.68\% &	2.76\%	\\
         w/o $\varphi$ &	 33.64\% &	0.32\% &	0.60\% &	1.16\%\\
         w/o $\mathcal{L}_q$ & 24.28\% & 	0.76\% & 	2.84\%	 & 3.28\%		\\
         
         \bottomrule
    \end{tabular}
    \caption{Ablation study for Anti-NAF algorithm on LAION-mi.}
    \label{tab:ablation_laion}
\end{table*}

